\documentclass[11pt,letterpaper]{article}

\usepackage{latexsym,multirow}
\usepackage{amssymb,amsmath, bm,mathtools}
\usepackage{graphicx}

\usepackage{enumerate}

\usepackage{natbib}
\usepackage[pdftex, bookmarksopen=true, bookmarksnumbered=true,
pdfstartview=FitH, breaklinks=true, urlbordercolor={0 1 0}, citebordercolor={0 0 1}]{hyperref}
\usepackage{colortbl}
\usepackage{subfigure}

\usepackage{dcolumn}
\newcolumntype{.}{D{.}{.}{-1}}
\newcolumntype{d}[1]{D{.}{.}{#1}}
\usepackage{theorem}
\theoremstyle{plain}
\theoremheaderfont{\scshape}
\newtheorem{assumption}{Assumption}

\newtheorem{theorem}{Theorem}

\newtheorem{lemma}{Lemma}
\newtheorem{condition}{Condition}

\usepackage{rotating}

\usepackage[compact]{titlesec}

\allowdisplaybreaks

\newcommand\spacingset[1]{\renewcommand{\baselinestretch}%
{#1}\small\normalsize}

\newcommand{\blind}{0}

\newcommand*{\QEDB}{\hfill\ensuremath{\square}}

\newcommand{\bone}{\mathbf{1}}

\newcommand{\pr}{\textnormal{pr}}
\newcommand{\var}{\textnormal{var}}
\newcommand{\cov}{\textnormal{cov}}

\newcommand{\bZ}{\bm{Z}}

\newcommand{\bz}{\bm{z}}

\newcommand{\oY}{\overline{Y}}
\newcommand{\wY}{\widehat{Y}}

\newcommand{\bA}{\bm{A}}
\newcommand{\ba}{\bm{a}}
\newcommand{\bY}{\bm{Y}}
\newcommand{\E}{\mathbb{E}}
\newcommand{\bX}{\mathbf{X}}
\newcommand{\bI}{\mathbf{I}}

\newcommand{\cM}{\mathcal{M}}
\newcommand{\bP}{\mathbf{P}}

\newcommand{\bW}{\mathbf{W}}
\newcommand{\ATE}{\textsc{ATE}}
\newcommand{\bepsilon}{\bm{\epsilon}}

\newcommand{\ADE}{\textsc{ADE}}

\newcommand{\MDE}{\textsc{MDE}}
\newcommand{\ASE}{\textsc{ASE}}

\begin{document} 

\newcommand{\tit}{Statistical Inference and Power Analysis for Direct and Spillover Effects in Two-Stage Randomized Experiments}
%
%
\spacingset{1.25}

\if0\blind

{\title{{\bf\tit}\thanks{Imai thanks the Alfred P. Sloan
      Foundation for partial support (Grant number 2020--13946).}}

  \author{Zhichao Jiang\thanks{School of Mathematics, Sun Yat-sen University, Guangzhou,  Guangdong 510275, China.}  \and Kosuke Imai\thanks{Professor, Department of Government and
      Department of Statistics, Institute for Quantitative Social
      Science, Harvard University, Cambridge MA 02138, USA. Phone:
      617--384--6778, Email:
      \href{mailto:Imai@Harvard.Edu}{Imai@Harvard.Edu}, URL:
      \href{https://imai.fas.harvard.edu}{https://imai.fas.harvard.edu}} \and Anup Malani\thanks{University of Chicago Law School and Pritzker School of
      Medicine, Chicago IL 60637, U.S.A, National Bureau of Economic
      Research, Cambridge MA 02138, U.S.A.}
      }

\date{
\today
}

\maketitle

}\fi

\if1\blind
\title{\bf \tit}

\maketitle
\fi

\pdfbookmark[1]{Title Page}{Title Page}

\thispagestyle{empty}
\setcounter{page}{0}

\begin{abstract}
 Two-stage randomized experiments are becoming an increasingly
  popular experimental design for causal inference when the outcome of
  one unit may be affected by the treatment assignments of other units
  in the same cluster. In this paper, we provide a methodological
  framework for general tools of statistical inference and power
  analysis for two-stage randomized experiments.  Under the
    randomization-based framework, we consider the estimation of a new
    direct effect of interest as well as the average direct and
    spillover effects studied in the literature.  We provide unbiased
    estimators of these causal quantities and their conservative
    variance estimators in a general setting.  Using these results, we
    then develop hypothesis testing procedures and derive sample
    size formulas.  We theoretically compare the two-stage randomized
  design with the completely randomized and cluster randomized
  designs, which represent two limiting designs. Finally, we conduct
  simulation studies to evaluate the empirical performance of our
  sample size formulas. For empirical illustration, the proposed
  methodology is applied to the randomized
  evaluation of the Indian national health insurance program. An
  open-source software package is available for implementing the
  proposed methodology.
 
 \bigskip 

\noindent {\bf Keywords:} experimental design, interference between
units, partial interference, spillover effects, statistical power
\end{abstract}


\clearpage
\spacingset{1.83}

\section{Introduction}

Much of the early causal inference literature relied upon the
assumption that the outcome of one unit cannot be affected by the
treatment assignment of another unit. Over the last two decades,
however, researchers have made substantial progress by developing a
variety of methodological tools to relax this assumption
 \citep[e.g.,][]{
  hudgens2008toward,tche:vand:12,
  forastiere2016identification, aron:sami:17,
  imai:jian:mala:21}.

Two-stage randomized experiments have become increasingly popular when
studying spillover effects. Under this experimental design,
researchers first randomly assign clusters of units to different
treatment assignment mechanisms, each of which has a different
probability of treatment assignment. For example, one treatment
assignment mechanism may randomly assign 80\% of units to the
treatment group whereas another mechanism may only treat 40\%.  Then,
within each cluster, units are randomized to the treatment and control
conditions according to its selected treatment assignment
mechanism. By comparing units who are assigned to the same treatment
conditions but belong to different clusters with different treatment
assignment mechanisms, one can infer how the treatment conditions of
other units within the same cluster affect one's outcome. Two-stage
randomized experiments are now frequently used in a number of
disciplines, including economics
\citep[e.g.,][]{ange:dima:16}, 
education \citep[e.g.,][]{roge:fell:18}, 
political science \citep[e.g.,][]{sinc:mcco:gree:12}, and public
health \citep[e.g.,][]{benj:etal:18}.

The increasing use of two-stage randomized experiments in applied
scientific research calls for the development of a general methodology
for analyzing and designing such experiments. Building on the prior
literature
\citep[e.g.,][]{hudgens2008toward,basse2016analyzing,imai:jian:mala:21},
we consider various direct and spillover effects, and develop their
unbiased point estimators and conservative variance estimators under
the nonparametric randomization-based framework.  This framework has
also been used to study other types of randomized designs
\citep[e.g.,][]{balzer2015adaptive,balzer2016targeted}. We also show
how to conduct hypothesis tests and derive the sample size formulas
for the estimation of these causal effects. The resulting formulas can
be used to conduct power analysis when designing two-stage randomized
experiments. Finally, we theoretically compare the two-stage
randomized design with its two limiting designs, the completely
randomized and cluster randomized designs. Through this comparison, we
analyze the potential efficiency loss of the two-stage randomized
design when no spillover effect exists.

We make several methodological contributions.  First, the proposed
causal quantities generalize those of \citet{hudgens2008toward} to
more than two treatment assignment mechanisms. We consider the joint
estimation of the average direct and spillover effects to characterize
the causal heterogeneity across different treatment assignment
mechanisms. We also propose the average marginal direct effect as a
scalar summary of several average direct effects. Second, our variance
estimators are guaranteed to be conservative while those of
\citet{hudgens2008toward} are not when applied to our setting.
Third, we develop hypothesis testing procedures and sample size
formulas, which can be used when planning a two-stage randomized
experiment.  Fourth, we prove the equivalence relationships between
the proposed randomization-based estimators and the least squares
estimators. These results extend those of \citet{basse2016analyzing},
in which the clusters have at most one treated unit.  Finally, an
open-source software package is available for implementing the
proposed methodology \citep{huan:jian:imai:22}.

In a closely related article, \citet{bair:etal:18} adopt a super
population framework to study the randomized saturation design (a
general form of two-stage randomized experiments), in which the
proportion of treated units for each cluster is drawn from a
distribution. The authors consider the assumptions about the structure
of spillover effects that are similar to those made in this
paper.  However, \citeauthor{bair:etal:18} impose a specific
  variance-covariance structure for potential outcomes and derive the
  standard errors of the causal estimates from a saturated linear
  model.  In contrast, we adopt the nonparametric randomization-based
  framework without imposing any variance-covariance structure for the
  potential outcomes although we consider simplifying conditions to
  facilitate the use of our method in practice. In addition, while
their goal is to determine the optimal distribution of the treated
proportion, we treat this distribution to be fixed and focus on the
development of estimators, hypothesis testing procedures, and sample
size formulas.

The remainder of the paper is organized as follows.
Section~\ref{sec::motivation} introduces our motivating study
concerning the impact evaluation of the Indian national health
insurance program \citep{imai:jian:mala:21,malanietal2021}.
Section~\ref{sec::design} formally presents the two-stage randomized
design and defines the three causal quantities of interest. In
Section~\ref{sec::general}, we propose a methodology for statistical
inference and power analysis.
Section~\ref{sec::application} revisits the health insurance study and
applies the proposed methods.  Finally, Section~\ref{sec::discussion}
provides concluding remarks.  The appendix presents simulation
studies, establishes the equivalence relations between the
regression-based and randomization-based inference, and compares the
two-stage randomized design with the cluster and individual randomized
designs.  All proofs appear in the Web Appendix.

\section{Randomized Evaluation of the Indian Health Insurance Program}
\label{sec::motivation}

We describe the randomized evaluation of the Indian national health
insurance program, which serves as our motivating application. In
2008, the Indian government introduced its first national public
health insurance scheme, Rastriya Swasthya Bima Yojana (RSBY). The
goal was to provide insurance coverage for hospitalization to
households below the poverty line.  Subsequently, the government
considered the expansion of the RSBY to some households above the
poverty line.

We conducted a randomized control trial to assess whether the
expansion of the RSBY increases access to hospitalization, and thus
health. The experiment took place in two districts of Karnataka State,
Gulbarga and Mysore.  Gulbarge has a total of 918 villages with the
village size varying from 0 to 2,428, while Mysore has 1,336 villages
with the size ranging from 0 to 2,976. We selected 22\% and 16\% of
the villages in Gulbarga and Mysore, respectively. This led to 11,089
households who had no pre-existing health insurance coverage and lived
within 25 km of an RSBY empaneled hospital.  The households in the
treatment group were offered an opportunity to enroll in the RSBY,
whereas those in the control group were able to buy the RSBY at the
usual government price.


\begin{table}
 \caption{The Two-stage Randomized Design for the Evaluation of the
Indian Health Insurance Program.} 
\begin{center}
 \begin{tabular}{lccc}
 & \multicolumn{3}{c}{Treatment assignment mechanisms} \\ \hline
 & 1 & 2 & 3  \\ \hline
 Treatment assignment proportion &  90\% & 70\% & 50\% \\
 Number of villages & 285 & 88 & 63 \\
 Number of households & 5512 & 1553& 1170 \\  
  \hline
 \end{tabular}
 \end{center}
\label{tb:design}
\end{table}

The evaluation was conducted using the two-stage randomized design
shown in Table~\ref{tb:design}.  In the first stage, a total of
  436 villages are randomly assigned to three treatment assignment
  mechanisms, yielding 258, 88, and 63 villages for treatment
  assignment mechanisms 1, 2, and 3, respectively.  Treatment
  assignment mechanisms 1, 2, and 3, correspond to the treatment
  assignment probabilities of 90\%, 70\%, and 50\%, respectively. In
the second stage of randomization, households were assigned to the
treatment within each village according to the treatment assignment
probability chosen in the first stage.  Households were informed of
the opportunities to enroll in RSBY from April to May,
2015. Approximately 18 months later, we carried out a survey and
measured a variety of outcomes about the health and financial
conditions of the household members. For more details about the
experiment, see
\citet{imai:jian:mala:21}, and \citet{malanietal2021}.

Both direct and spillover effects are of interest.  The direct effect
quantifies how much the household members would benefit from their own
receipt of the program benefits. In contrast, the spillover effect
characterizes how the treatment of other households affects one's
outcomes, possibly through the replacement of informal insurance by
formal insurance and the efficient use of limited resources in local
hospitals. Moreover, the heterogeneity in the direct and spillover
effects is also of interest. For example, a greater treatment
assignment probability may cause the overcrowding of local hospitals,
leading to a lower direct effect.



\section{Experimental Design and Causal Quantities of Interest}
\label{sec::design}

We now formally describe the two-stage randomized experimental design
and define the causal quantities of interest using the potential
outcomes framework \citep[e.g.,][]{neym:23,rubi:74a}.

\subsection{Assumptions}

Suppose that we have a total of $J$ clusters and each cluster $j$ has
$n_j$ units. Let $N$ represent the total number of units, i.e.,
$N = \sum_{j=1}^J n_j$. Under the two-stage randomized design, we
first randomly assign clusters to different treatment assignment
mechanisms, and then assign a certain proportion of individual units
within a cluster to the treatment condition by following the treatment
assignment mechanism selected at the first stage of randomization.
Let $A_j$ denote the treatment assignment mechanism chosen for cluster
$j$, which takes a value in $\cM=\{1, 2, \ldots, m\}$. Let
$\bA= (A_1, A_2,\dots, A_J)$ denote the vector of treatment assignment
mechanisms for all $J$ clusters and $\ba=(a_1,a_2,\ldots,a_J)$
represent the vector of realized assignment mechanisms. We assume
complete randomization such that a total of $J_a$ clusters are
assigned to the assignment mechanism $a \in \cM$ where
$ \sum_{a=1}^{m} J_a = J$.

The second stage of randomization concerns the treatment assignment
for each unit within cluster $j$ based on the assignment mechanism
$A_j$. Let $Z_{ij}$ be the binary treatment assignment variable for
unit $i$ in cluster $j$ where $Z_{ij}=1$ and $Z_{ij} = 0$ imply that
the unit is assigned to the treatment and control conditions,
respectively. Let
$\bZ_j=(Z_{1j}, \ldots,Z_{n_j j})$ be the vector of assigned
treatments for the $n_j$ units in the cluster and
$\bz_j=(z_{1j},\ldots,z_{n_j j})$ be the vector of realized
assignments.
 Then, $\Pr(\bZ_j = \bz_j \mid A_j = a)$ represents the
distribution of the treatment assignment when cluster $j$ is assigned
to the assignment mechanism $A_j = a$. We assume complete randomization such that a total of
$n_{jz}$ units in cluster $j$ are assigned to the treatment condition
$z \in \{0,1\}$ where $n_{j0}+n_{j1}=n_j$. 
Finally, let $\bZ=(\bZ_1,\ldots,\bZ_J)$ be the vector of assigned treatments for all the $N$ units in the population and $\bz=(\bz_1,\ldots,\bz_J)$ be the vector of realized assignments.
We now formally define the
two-stage randomized design.

\begin{assumption} {\sc (Two-Stage Randomization)} \label{asm::two-stage}
 \begin{enumerate} 
 \item Complete randomization of treatment assignment mechanisms
 across clusters:
 \begin{equation*}
 \Pr(\bA = \ba) \ = \ \frac{J_1!\cdots J_m!}{J!}
 \end{equation*}
 for all $\ba$ such that
 $\sum_{j=1}^J \bone(a_j=a^\prime) = J_{a^\prime}$ for
 $a^\prime \in \cM$.
 \item Complete randomization of treatment assignment across units within each
 cluster:
 \begin{equation*}
 \Pr(\bZ_j = \bz_j \mid A_j = a) \ = \ \frac{1}{\binom{n_j}{n_{j1}}}
 \end{equation*}
 for all $\bz_j$ such that $\sum_{i=1}^{n_j} z_{ij}= n_{j1}$.
 \end{enumerate}
\end{assumption}

Next, we introduce the potential outcomes. For unit $i$ in cluster
$j$, let $Y_{ij}(\bz)$ be the potential value of the outcome if the
assigned treatment vector for the entire sample is $\bz$ where $\bz$
is an $N$ dimensional vector. The observed outcome is given by
$Y_{ij}=Y_{ij}(\bZ)$. This notation implies that the outcome of one
unit may be affected by the treatment assignment of any other unit in
the sample.

Unfortunately, it is impossible to learn about causal effects without
additional assumptions because each unit has $2^N$ possible potential
outcome values.  Thus, following the literature
\citep{sobel2006randomized,hudgens2008toward}, we assume that the
potential outcome of one unit cannot be affected by the treatment
assignment of another unit in other clusters while allowing for
possible interference between units within a cluster.
\begin{assumption}[No Interference Between Clusters]
\label{asm::nointer}
$$Y_{ij}(\bz) \ = \ Y_{ij}(\bz') \
 {\rm for\ any} \ \bz, \bz^\prime \ {\rm with} \ \bz_j \ = \ \bz'_j.$$
\end{assumption}
Assumption~\ref{asm::nointer}, which is known as the partial
interference assumption in the literature, partially relaxes the
standard assumption of no interference between units \citep{rubi:90}.
This assumption reduces the number of potential outcome values for
each unit from $2^N$ to $2^{n_j}$. 

Lastly, we rely upon the stratified interference assumption proposed
by \citet{hudgens2008toward} to further reduce the number of potential
outcome values.
\begin{assumption}[Stratified Interference]
\label{asm::stratified}
\begin{equation*}
Y_{ij}(\bz_j) \ = \ Y_{ij}(\bz^\prime_j) \quad {\rm if} \quad z_{ij}=z^\prime_{ij}
\ {\rm and} \ \sum_{i=1}^{n_j} z_{ij} \ = \ \sum_{i=1}^{n_j} z^\prime_{ij}.
\end{equation*}
\end{assumption}
Assumption~\ref{asm::stratified} implies that the outcome of one unit
depends on the treatment assignment of other units only through the
number of those who are assigned to the treatment condition within the
same cluster.  The assumption has been commonly used in the literature
\citep[e.g.,][]{tche:vand:12,liu2014large,miles2019causal}.  It is a
reasonable simplification of the interference structure and is
directly motivated by two-stage randomization which varies the
proportion of treated units within a cluster.

As pointed out by \citet{hudgens2008toward}, although the
identification of the direct and indirect effects does not require
Assumption~\ref{asm::stratified}, a valid variance estimator is
unavailable without an additional assumption.  A more general form of
Assumption~\ref{asm::stratified} is exposure mappings, which require
the potential outcome to depend on a known function of treatment
conditions
\citep[e.g.,][]{vanderweele2013mediation,forastiere2016identification,bargagli2020heterogeneous,forastiere2021identification,savje2021average}.
While it is relatively straightforward to extend our results regarding
statistical inference under such settings \citep{aron:sami:17}, sample
size and power calculation will be more complicated.  Therefore, we
maintain Assumption~\ref{asm::stratified} throughout this paper.
Under Assumptions~\ref{asm::nointer}~and~\ref{asm::stratified}, we can
simplify the potential outcome as a function of one's own treatment
and the treatment assignment mechanism of its cluster, i.e.,
$Y_{ij}(\bz)=Y_{ij}(z,a)$.

\subsection{Direct effect}

Under the above assumptions, we now define the main causal quantities
of interest. The first quantity is the direct effect of the treatment
on one's own outcome. We define the unit-level direct effect for unit
$i$ in cluster $j$ as,
\begin{eqnarray*}
 \ADE_{ij}(a) & = & Y_{ij}(1,a)-Y_{ij}(0,a)
\end{eqnarray*}
for $a=1,\ldots,m$. This quantity may depend on the treatment
assignment mechanism $a$ due to the possible spillover effect from
other units' treatments. The direct effect quantifies how the
treatment of a unit affects its outcome under a specific assignment
mechanism. This unit-level direct effect can be aggregated, leading
to the definition of the cluster-level direct effect,
\begin{eqnarray*}
\ADE_{j}(a) \ = \ \frac{1}{n_j} \sum_{i=1}^{n_j} \ADE_{ij}(a) \ = \ \oY_j(1,a)-\oY_j(0,a),
\end{eqnarray*}
where
$ \oY_j(z,a) \ = \ 1/n_j\cdot \sum_{i=1}^{n_j} Y_{ij}(z,a)$.
We can further aggregate this quantity and obtain the population-level 
direct effect,
\begin{equation}
\label{eqn::ade}
 \ADE(a) \ = \ \frac{1}{J} \sum_{j=1}^J \ADE_j(a)=\oY(1,a)-\oY(0,a),
\end{equation}
where  $\oY(z,a) \ = \ 1/J \cdot \sum_{j=1}^J \oY_j(z,a)$.
The direct effects depend on the treatment assignment mechanisms; we
denote them by a column vector,
$\ADE =(\ADE(1),\ldots, \ADE(m))^\top$.

\subsection{Marginal direct effect}

With $m$ treatment assignment mechanisms, we have a total of $m$
direct effects $\ADE(a)$ for $a=1,\ldots,m$. Although such direct
effects are informative about how the treatment of a unit affects its
own outcome given different treatment assignment mechanisms,
researchers may be interested in having a single quantity that
summarizes all the direct effects. We define the unit-level marginal
direct effect by marginalizing the direct effects over the treatment
assignment mechanisms,
\begin{eqnarray*}
 \MDE_{ij} & = & \sum_{a=1}^m q_a \{ Y_{ij}(1,a)-Y_{ij}(0,a)\}.
\end{eqnarray*}
The weight $q_a$ is the proportion of the clusters assigned to
treatment assignment mechanism $a$, which equals  $J_a/J$ under Assumption~\ref{asm::two-stage}. Based on the unit-level effect,
we define the cluster-level marginal direct effect and the population-level marginal direct effect as
\begin{equation}
\label{eqn::mde}
\MDE_{j}  =  \frac{1}{n_j} \sum_{i=1}^{n_j} \MDE_{ij},\quad \MDE \ = \ \frac{1}{J} \sum_{j=1}^J\MDE_j.
\end{equation}
We emphasize that the MDE, unlike the ADE, depends on the distribution
of treatment assignment mechanism $q_a$.  Thus, a different value of
the design parameter can alter the interpretation of  MDE.

\subsection{Spillover effect}

In two-stage randomized experiments, another causal quantity of
interest is the spillover effect, which quantifies how one's treatment
affects the outcome of another unit. Under
Assumptions~\ref{asm::nointer}~and~\ref{asm::stratified}, we define
the unit-level spillover effect on the outcome as,
\begin{eqnarray*}
 \ASE_{ij}(z;a,a') \ = \ Y_{ij}(z,a)-Y_{ij}(z,a'), 
\end{eqnarray*}
which compares the potential outcomes under two different
assignment mechanisms, $a $ and $a'$, while holding one's treatment
assignment constant at $z$. We then define the spillover effects on
the outcome at the cluster and population levels,
\begin{eqnarray*}
 \ASE_{j}(z;a,a') & = & \frac{1}{n_j} \sum_{i=1}^{n_j}
 \ASE_{ij}(z;a,a'),\quad
 \ASE(z;a,a') \ = \ \frac{1}{J} \sum_{j=1}^J \ASE_{j}(z;a,a').
\end{eqnarray*}
The spillover effects depend on both the treatment condition and
treatment assignment mechanisms; we denote them by
$\ASE =
(\ASE(1;1,2),\ASE(1;2,3),\ldots,\ASE(1;m-1,m),$ $\ASE(0;1,2),\ASE(0;2,3),\allowbreak\ldots,\ASE(0;m-1,m))$,
which consists of the spillover effects comparing adjacent treatment
assignment mechanisms for both the treatment and control conditions.

We give equal weight to each cluster in the quantities defined above
\citep[see][]{hudgens2008toward}, while \citet{basse2016analyzing}
assign an equal weight to each unit. For example,
\citet{basse2016analyzing} define the direct effect as
\begin{eqnarray*}
 \ADE(a) & = & \sum_{j=1}^J \frac{n_j}{N}\cdot \ADE_{j}(a) \ = \ \frac{1}{N} \sum_{j=1}^J\sum_{i=1}^{n_j} \ADE_{ij}(a).
\end{eqnarray*}
While our analysis focuses on the cluster-weighted quantities rather
than individual-weighted quantities, our method can be generalized to
any weighting scheme.

\section{A General Methodology for Two-Stage Randomized Experiments}
\label{sec::general}

We next develop a general methodology for the direct and spillover
effects introduced above. We show how to estimate these quantities,
compute the randomization-based variance, and conduct hypothesis
tests. We also derive the sample size formulas for testing the direct
and spillover effects.

Formally, define
$\oY = (\oY(1,1),\oY(0,1), \ldots,\oY(1,m),\oY(0,m))^\top$, which is a
$2m$-dimensional column vector with the $(2a-1)$-th and $2a$-th
elements representing the treatment and control potential outcomes
under treatment assignment mechanism $a$, respectively, for
$a=1,\ldots,m$. The direct, marginal direct, and spillover effects can
all be written as linear transformations of $\oY$.  Our methodological
development will exploit these linear transformations.

In particular, let $e_{l}$ denote the $2m$-dimensional column vector
whose $l$-th element is equal to $1$ with other elements being equal
to $0$. Then, the direct effect can be written as $\ADE = C_1 \oY$,
where
$C_1 = (e_{1}-e_{2},e_{3}-e_{4},\ldots, e_{2m-1}-e_{2m})^\top$
is an $m\times 2m$ matrix with the $a$-th row representing the
contrast in $\ADE(a)$ for $a=1,\ldots,m$. Similarly, the marginal
direct effect can be written as $\MDE = C_2 \oY$, where
$C_2=(J_1,-J_1,J_2,-J_2,\ldots,J_m,-J_m)^\top/J$. Lastly, the
spillover effect can be written as $\ASE =C_3 \oY$, where
$C_3 = (C_{31}, C_{30})^\top$ with
$C_{31} = (e_1-e_3, e_3-e_5,\ldots, e_{2m-3}-e_{2m-1})^\top$ and
$C_{30} = (e_2-e_4, e_4 -e_6,\ldots, e_{2m-2}-e_{2m})^\top$. That is,
the $a$-th column in $C_{31}$ and $C_{30}$ represents the contrast in
$\ASE(1;a,a+1)$ and $\ASE(0;a,a+1)$, respectively, for
$a=1,\ldots,m-1$. 

Under Assumptions~\ref{asm::nointer}~and~\ref{asm::stratified}, our
setting is similar to a split-plot design in the sense that the
treatment and the treatment assignment mechanism can be viewed as the
interventions at the sub-plot and whole-plot levels,
respectively. Therefore, we leverage this connection and use the
results in the split-plot design developed in
\citet{zhao2021reconciling} to obtain the unbiased estimation,
variances of the estimators, and asymptotic properties of the
estimators. We then develop hypothesis testing procedures and sample
size formulas based on these results.
 
\subsection{Unbiased estimation}

\citet{hudgens2008toward} propose unbiased estimators of the average
direct and spillover effects. Here, we present analogous estimators
for the three causal quantities defined above. Define
 \begin{eqnarray*}\widehat{Y}_j(z)\ = \ \frac{\sum_{i=1}^{n_j} Y_{ij}
 \bone(Z_{ij}=z)}{\sum_{i=1}^{n_j} \bone(Z_{ij}=z)} \quad {\rm and} \quad
 \widehat{Y}(z,a)\ = \ \frac{
 \sum_{j=1}^{J}\widehat{Y}_j(z)
 \bone(A_{j}=a)}{\sum_{j=1}^{J} \bone(A_{j}=a)},
 \end{eqnarray*}
 where $\widehat{Y}_j(z)$ is the average outcome under treatment
 condition $z$ in cluster $j$, and $\widehat{Y}(z,a)$ is the average
 of $\widehat{Y}_j(z)$ in clusters with treatment assignment mechanism
 $a$. The following theorem gives the unbiased estimators of the \ADE,
 \MDE, and \ASE.

\begin{theorem}[Unbiased Estimation]
\label{thm::identification}  
Define $\wY = (\wY(1,1),\wY(0,1), \ldots,\wY(1,m),\wY(0,m))$. Under
Assumptions~\ref{asm::two-stage},~\ref{asm::nointer},~and~\ref{asm::stratified},
$\wY$ is unbiased for $\oY$, i.e., $\E(\wY) = \oY$. Therefore,
$\widehat \ADE=C_1\wY$, $\widehat \MDE=C_2\wY$, and
$\widehat \ASE=C_3\wY$ are unbiased for $\ADE$, $\MDE$, and $\ASE$,
respectively, i.e.,
$\E(\widehat \ADE) \ = \ \ADE, \quad \E(\widehat \MDE) \ = \ \MDE,
\quad \E(\widehat \ASE) \ = \ \ASE.$
\end{theorem}
We note that the theory of simple random sampling implies
$\E\{\wY_j(z) \mid A_j=a \} = \oY_j(z,a)$. Therefore, it is
straightforward to show that $\E\{\wY(z,a)\}= \oY(z,a)$ and hence
$\E(\wY)=\oY$.

\subsection{Variance}

\citet{hudgens2008toward} derive the variances of $\widehat{\ADE}(a)$
and $\widehat{\ASE}(z;a',a)$ under stratified interference
(Assumption~\ref{asm::stratified}). However, this is not sufficient
for obtaining the variance of our causal quantities, which require the
covariance between the elements in $\wY$. We first derive the
covariance matrix of $\wY$ and then use it to obtain the covariance
matrix of \ADE, \MDE, and \ASE.

The covariance matrix of $\wY$ consists of the variance of $\wY(z,a)$
and the covariance between $\wY(z,a)$ and $\wY(z',a')$. Define,
\begin{eqnarray*}
\sigma_j^2(z,z';a,a') &=& \frac{1}{n_j-1} \sum_{i=1}^{n_j} \{Y_{ij}(z,a)-
 \oY_{j}(z,a)\}\{Y_{ij}(z',a')- \oY_{j}(z',a')\},\\
 \sigma_b^2(z,z';a,a') &=& \frac{1}{J-1} \sum_{j=1}^{J} \{\oY_j(z,a)-
 \oY(z,a)\}\{\oY_j(z',a')- \oY(z',a')\},
\end{eqnarray*}
where $\sigma_j^2(z,z';a,a')$ is the within-cluster covariance between
$Y_{ij}(z,a)$ and $Y_{ij}(z',a')$, and $\sigma_b^2(z,z';a,a')$ is
their between-cluster covariance.  When $a=a'$,
$\sigma_j^2(z,z';a,a')$ reduces to $\sigma_j^2(z,z';a)$ and
$\sigma_b^2(z,z';a,a')$ equals $\sigma_b^2(z,z';a)$. When $z=z'$,
$\sigma_j^2(z,z';a,a')$ reduces to $\sigma_j^2(z;a,a')$ and
$\sigma_b^2(z,z';a,a')$ equals $\sigma_b^2(z;a,a')$.  Lastly, when
$z=z'$ and $a=a'$, $\sigma_j^2(z,z';a,a')$ reduces to
$\sigma_j^2(z,a)$ and $\sigma_b^2(z,z';a,a')$ equals
$\sigma_b^2(z,a)$. We denote
$S_b = (\sigma^2_b(z,z';a,a') )_{2m \times 2m}$ and
$S_j = (\sigma^2_j(z,z';a,a') )_{2m \times 2m}$ as the between and
within cluster covariance matrix of
$(Y_{ij}(1,1),Y_{ij}(0,1),\ldots,Y_{ij}(1,m),Y_{ij}(0,m))$.

Let $0_{m\times n }$ and $1_{m\times n }$ be the $m\times n$ matrices
of zeros and ones, respectively, whereas $I_m$ is the $m\times m$
identity matrix.  Use $\otimes$ and $\circ$ to denote the Kronecker
and Hadamard products of matrices, respectively. Denote
\begin{eqnarray*}
H & = & \text{diag}(J/J_1,\ldots,J/J_m ) \otimes 1_{2\times 2} - 1_{2m\times 2m},\\
H_j &=& \text{diag}(J/J_1,\ldots,J/J_m ) \otimes \left\{ \text{diag}(n_j/n_{j1},n_j/n_{j0} )-1_{2\times 2} \right\}.
\end{eqnarray*}

The following theorem gives the covariance matrix of
$\wY$.
\vspace{-0.2cm}
\begin{theorem}[Variance-Covariance Matrix]
 \label{thm::var-general} Under
 Assumptions~\ref{asm::two-stage},~\ref{asm::nointer},~and~\ref{asm::stratified},
 we have
\begin{eqnarray*}
\cov(\wY)\ = \ J^{-1} (H \circ S_b)+J^{-2} \sum_{j=1}^J n_j^{-1} ( H_j \circ S_j).
\end{eqnarray*}
\end{theorem}
The multiplication facilitates the development of sample size formulas
in Section~\ref{subsec:sample_size}.  Theorem~\ref{thm::var-general}
implies that the covariance matrices of $\widehat{\ADE}$,
$\widehat{\MDE}$, and $\widehat{\ASE}$ are
$ \var\{\widehat \ADE\} \ = \ C_1 DC_1^\top/J,\quad \var\{\widehat
\MDE\} \ = \ C_2 DC_2^\top/J,\quad \var\{\widehat \ASE\} \ = \ C_3
DC_3^\top/J $ where $D=J \cov(\wY)$.

Because we cannot observe $Y_{ij}(1,a)$ and $Y_{ij}(0,a)$
simultaneously, no unbiased estimator exists for
$\sigma^2_j(1,0;a)$.  This implies that no
unbiased estimation of $D$ is possible. Following the idea of
\citet{hudgens2008toward}, we propose a conservative estimator.
Define
\begin{eqnarray*}
 \widehat \sigma_b^2(z,a)&=&\frac{1}{J_a-1}\sum_{j=1}^{J} \left \{ \wY_j(z)- \wY(z,a)\right \}^2\bone(A_j=a),\\
 \widehat \sigma_b^2(1,0;a)&=&\frac{1}{J_a-1}\sum_{j=1}^{J} \left \{ \wY_j(1)- \wY(1,a)\right \}\left \{ \wY_j(0)- \wY(0,a)\right \}\bone(A_j=a),
\end{eqnarray*}
where $\widehat{\sigma}_b^2(z,a)$ represents the between-cluster
sample variance of $Y_{ij}(z,a)$, and $\widehat{\sigma}_b^2(1,0;a)$
denotes the between-cluster sample covariance between $Y_{ij}(1,a)$
and $Y_{ij}(0,a)$. The following theorem provides a conservative
variance estimator, which is exactly unbiased when the cluster-level
average potential outcome, i.e., $\oY_j(z,a)$, does not vary across
clusters.
\begin{theorem}[Conservative Estimator of Variance]
\label{thm::varest}
Let $\widehat D$ be a $2m$ by $2m$ block diagonal matrix
with the $a$-th matrix ($a=1,\ldots,m$) on the diagonal 
$$
\widehat D_a \ = \ \frac{J}{J_a}\begin{pmatrix}
 \widehat \sigma^2_b(1,a) & \widehat \sigma^2_b(1,0;a) \\
 \widehat \sigma^2_b(1,0;a) & \widehat \sigma^2_b(0,a)
\end{pmatrix}.
$$ 
Then, $\widehat D$ is a conservative estimator for $D$,
i.e., $\E\{\widehat D\} - D$ is a positive semi-definite matrix. It is
an unbiased estimator for $D$ when the cluster-level average potential
outcomes, i.e., $\oY_j(z,a)$, is constant across clusters.
\end{theorem}
The covariance matrix estimator $\widehat D$ estimates
$\var \left\{\wY(z,a)\right\} $ and
$\cov \left\{\wY(1,a),\wY(0,a)\right\} $ by their corresponding
between-cluster sample variance and covariance, $ \widehat{\sigma}^2_b(z,a)$ and
$ \widehat{\sigma}^2_b(1,0;a)$, while replacing
$\cov\left\{\wY(1,a),\wY(0,a')\right\} $ with $0$.
Theorem~\ref{thm::varest} implies the following conservative variance
estimators for \ADE, \MDE, and \ASE,
$\widehat \var\{\widehat \ADE\} \ = \ C_1 \widehat
 DC_1^\top/J,\quad \widehat \var\{\widehat \MDE\} \ = \ C_2
 \widehat DC_2^\top/J,\quad \widehat \var\{\widehat \ASE\} \ = \ C_3\widehat DC_3^\top/J.$
Similar to $\widehat D$, these estimators are unbiased if $\oY_j(z,a)$ are the same across clusters.

Note that alternative conservative variance estimators exist with
different conditions for unbiasedness \citep{mukerjee2018using}. In
particular, \citet{hudgens2008toward} propose the following
conservative variance estimator for each $\ADE(a)$,
\begin{eqnarray*}
\frac{1}{J_a}\left( 1-\frac{J_a}{J}\right) \left\{\widehat \sigma^2_b(1,a)+\widehat \sigma^2_b(0,a)-2\widehat \sigma^2_b(1,0;a) \right\}+ \frac{1}{JJ_a} \sum_{j=1}^J \left\{\frac{\widehat\sigma^2_j(1)}{n_{j1}}+\frac{\widehat\sigma^2_j(0)}{n_{j0}}\right\} \bone(A_j=a),
\end{eqnarray*}
where $\widehat \sigma_j^2(z) \ = \ 1/(n_{jz}-1)\cdot \sum_{i=1}^{n_J} \{Y_{ij}- \wY_j(z)\}^2\bone(Z_{ij}=z)$
represents the within-cluster sample variance of $Y_{ij}(z)$. They
show that it is a conservative estimator of the variance of
$\ADE(a)$, and is unbiased if the unit-level direct effects,
$Y_{ij}(1,a)-Y_{ij}(0,a)$, do not vary within each cluster.

In practice, this variance estimator is generally smaller than the
$a$-th diagonal element of $\widehat \var\{\widehat \ADE\}$. However,
its conservativeness property holds only for the variance of each
$\ADE(a)$. No similar estimator can be obtained for the covariance
matrix of $\widehat \ADE$. For example, replacing the diagonal
elements of $\widehat \var\{\widehat \ADE\}$ with
\citet{hudgens2008toward}'s estimators does not yield a conservative
estimator for $\var\{\widehat \ADE\}$. Therefore, we recommend using
\citet{hudgens2008toward}'s estimator when the variance of $\ADE(a)$
alone is of interest whereas our proposed estimator should be used
when the joint distribution of $\ADE$ is of interest.

\subsection{Asymptotic normality of the estimators}

To conduct statistical inference and power analysis, we study the
asymptotic properties of the estimators. We state the regularity
conditions for finite-population asymptotics.
\begin{condition} 
\label{con::regularity}
Denote $\overline{Y^4_j(z,a)} = n_j^{-1} \sum_{i=1}^{nj} Y^4_{ij}(z,a)$. As $J$ goes to infinity, 
for $z=0,1$ and $a=1,\ldots,m$,
\begin{enumerate}[(a)]
\item $J_a/J$ has a limit in $(0,1)$; $\epsilon \leq n_{jz}/n_j \leq 1-\epsilon$ for $j=1,\ldots,J$, and some $\epsilon \in (0,1/2)$;
\item $\max_{j} |\oY_j(z,a)-\oY(z,a)|^2/J = o(1)$;
\item $\oY$ has a finite limit; $S_b = O(1)$ and $ J^{-1}\sum_{j=1}^J n_j^{-1} \{H_j \circ S_j\} = O(1)$;
\item $J^{-2} \sum_{j=1}^J \overline{Y^4_j(z,a)} = o(1)$.
\end{enumerate}
\end{condition}
From Theorem~\ref{thm::var-general}, 
Conditions~\ref{con::regularity}(a) and (b) imply that the covariance matrix of $\wY$ is at the order of $J^{-1}$, which guarantees the consistency of $\wY$ for estimating $\oY$. Conditions~\ref{con::regularity}(c) and (d) hold as long as $Y_i$ is bounded. Condition~\ref{con::regularity} requires only $J$ to go to infinity and thus can incorporate both scenarios when the cluster size is fixed or goes to infinity.
\vspace{-0.2cm}
\begin{theorem}[Asymptotic normality]
\label{thm::CLT}
 Under Assumptions~\ref{asm::two-stage},~\ref{asm::nointer},~\ref{asm::stratified},~and Condition~\ref{con::regularity}, we have 
$\sqrt{J}(\wY - \oY) \stackrel{d}{\rightarrow} N(0,D^\ast)$,
where $D^\ast$ is the limiting value of $D$.
\end{theorem}
\subsection{Hypothesis testing}
\label{subsec:hypothesis_testing}

We consider testing the following three null hypotheses of no direct
effect, no marginal direct effect, and no spillover effect, 
$H_0^{\text{de}}: \ADE = 0, \quad H_0^{\text{mde}}: \MDE = 0, \quad H_0^{\text{se}}: \ASE = 0.$
Because \ADE, \MDE, and \ASE{} are linear transformations of $\oY$, we
focus on a more general null hypothesis,
\begin{equation}
 H_0: C\oY = 0, \label{eqn::null-hyp} 
\end{equation}
where $C$ is a constant contrast matrix with full row rank. By setting
$C$ to $C_1$, $C_2$, and $C_3$, $H_0$ becomes $H_0^{\text{mde}}$,
$H_0^{\text{se}}$, and $H_0^{\text{se}}$, respectively.  We propose
the following Wald-type test statistic,
\begin{eqnarray}
\label{eqn::wald-stat}
T \ = \ J(C\wY)^\top (C\widehat DC^\top)^{-1}(C\wY),
\end{eqnarray}
where the covariance matrix of $\wY$ is replaced with its conservative
estimator $\widehat D/J$. Unfortunately, $T$ does not follow a
$\chi^2$ distribution asymptotically with the conservative covariance matrix
estimator. 
\vspace{-0.3cm}
\begin{theorem}[Asymptotic Distribution of the Test Statistic]
 \label{thm::test}Suppose that Assumptions~\ref{asm::two-stage},~\ref{asm::nointer},~\ref{asm::stratified},~and Condition~\ref{con::regularity} hold, and the rank of $C$ is $k$. Under the null hypothesis in Eqn.~\eqref{eqn::null-hyp},
 the asymptotic distribution of the test statistic $T$ defined in
 Eqn.~\eqref{eqn::wald-stat} is stochastically dominated by the
 $\chi^2$ distribution with $k$ degrees of freedom, i.e.,
 $\Pr(T\geq t) \leq \Pr\{X\geq t\}$ for any constant $t$ where
 $X \sim \chi^2(k)$.
\end{theorem}
 With a pre-specified
significance level $\alpha$, we can reject $H_0$ if
$T > \chi^2_{1-\alpha}(k)$ where $\chi^2_{1-\alpha}(k)$ represents the
$(1-\alpha)$ quantile of the $\chi^2$ distribution with $k$ degrees of
freedom. Theorem~\ref{thm::test} implies that this rejection rule
controls the type I error asymptotically.
 
We can use the following three Wald-type test statistics for the
direct, marginal direct, and spillover effects, respectively,
\begin{eqnarray}
T_{\text{de}} &=& J(C_1\wY)^\top (C_1\widehat DC_1^\top)^{-1}(C_1\wY),
 \label{eqn::de-teststat} \\
T_{\text{mde}} &=& J(C_2\wY)^\top (C_2\widehat
  DC_2^\top)^{-1}(C_2\wY), \label{eqn::mde-teststat}\\
T_{\text{se}} &=& J(C_3\wY)^\top (C_3\widehat
  DC_3^\top)^{-1}(C_3\wY). \label{eqn::se-teststat}
\end{eqnarray}
Theorem~\ref{thm::test} implies that under the corresponding null
hypothesis, the asymptotic distributions of $T_{\text{de}}$,
$T_{\text{mde}}$, and $T_{\text{se}}$ are stochastically dominated by
a $\chi^2$ distribution with the degrees of freedom equal to $m$, one,
and $2(m-1)$, respectively.

\subsection{Sample size formula}
\label{subsec:sample_size}
When planning a two-stage randomized experiment, we may wish to
determine the sample size needed to detect a certain effect size with
a given statistical power ($1-\beta$) and a significance level
($\alpha$). The sample size depends on the number of clusters and
cluster sizes. In two-stage randomized experiments, however, the
cluster sizes are often fixed. Therefore, we derive the required
number of clusters of fixed sizes that ensures sufficient power to
detect a deviation from the null hypothesis.

\paragraph{General formulation.}
We begin by considering a general alternative hypothesis,
\begin{equation}
 H_1: C\oY = x, \label{eqn::alternative-hyp}
\end{equation}
where $C$ is a $k \times 2m$ matrix of full row rank ($k \leq 2m$) and $x$ is a
vector of constants. With the test statistic given in
Eqn.~\eqref{eqn::wald-stat}, the required number of clusters $J$
should satisfy
\begin{eqnarray}
\label{eqn::samplesize-general}
\pr\{J(C\wY)^\top (C\widehat DC^\top)^{-1}(C\wY) \geq
 \chi^2_{1-\alpha}(k)\mid C\oY =x\} \ \geq \ 1-\beta.
\end{eqnarray}
However, because $\widehat D$ is a conservative estimator for $D$,
$J(C\wY)^\top (C\widehat DC^\top)^{-1}(C\wY)$ follows a generalized
chi-square distribution instead of a standard chi-square distribution
asymptotically, rendering it difficult to directly solve
Eqn.~\eqref{eqn::samplesize-general} for $J$.

Fortunately, based on the properties of the generalized chi-square
distribution, the following theorem gives a conservative sample size
formula.
\begin{theorem}[General sample size formula]
 \label{thm::samplesize-general}
 Consider a statistical hypothesis test with level $\alpha$ where the
 null and alternative hypotheses are given in
 Eqn.~\eqref{eqn::null-hyp}~and~\eqref{eqn::alternative-hyp},
 respectively. We reject the null hypothesis if
 $T > \chi^2_{1-\alpha}(k)$ where the test statistic $T$ is defined
 in Eqn.~\eqref{eqn::wald-stat} and $k$ is the rank of $C$.
 Then, the number of clusters required for this hypothesis test to
 have the statistical power of $(1-\beta)$ is given by,
\begin{eqnarray*}
 J \ \geq \ \frac{ s^2( \chi^2_{1-\alpha}(k), 1-\beta,k) }{x^\top \{C \E(\widehat D)C^\top\}^{-1} x },
\end{eqnarray*}
where $s^2( q, 1-\beta,k)$ represents the non-centrality parameter of
the non-central $\chi^2$ distribution with $k$ degrees of freedom
whose $\beta$ quantile is equal to $q$.
\end{theorem}
In practice, we must compute $s^2( \chi^2_{1-\alpha}(k), 1-\beta,k) $
numerically. Based on Theorem~\ref{thm::samplesize-general}, we can
obtain the sample size formula for the direct, marginal direct, and
spillover effects by setting $k$ to $m$, one, and $2(m-1)$,
respectively.


\paragraph{Simplification}
The practical difficulty of the sample size formula in
Theorem~\ref{thm::samplesize-general} is that it requires the specification
of many parameters in $\E(\widehat D)$ and the value of vector $x$ in
the alternative hypothesis. Thus, we consider the further
simplification of the sample size formula to facilitate its application
by reducing the number of parameters to be specified by researchers.

\begin{assumption}[Simplification] \label{asm::simplification} We
  make the following simplifying assumptions:
\begin{enumerate}[(a)]
\item The within-cluster variances of $Y_{ij}(z,a)$ are the same
 across different clusters, different treatments, and different
 treatment assignment mechanisms: $\sigma_j^2(z,a) = \sigma^2_w$ for
 all $z, a$;
\item The between-cluster variances of $Y_{ij}(z,a)$ are the same
 across different treatments and different treatment assignment
 mechanisms: $\sigma_b^2(z,a) = \sigma^2_b$ for all $z$ and $a$;
\item The within-cluster and between-cluster correlation coefficients
 between $Y_{ij}(1,a)$ and $Y_{ij}(0,a)$ are the same and
 non-negative: $\sigma^2_j(1,0;a) =\sigma^2_{j'}(1,0;a') \geq 0 $ and
 $\sigma^2_b(1,0;a) =\sigma^2_b(1,0;a')\geq 0$ for all $j$, $j'$, $a$ and $a'$.
\end{enumerate}
\end{assumption} 
 \citet{bair:etal:18} also make simplifying
  assumptions to reduce the number of parameters. The authors,
  however, use the super population framework to derive the optimal
  design parameters rather than the sample size formulas as done in
  this paper. 

Under these simplifying conditions, we can write
$\sigma^2_j(1,0;a) = \rho \sigma_w^2 $ and
$\sigma^2_b(1,0;a) = \rho \sigma_b^2 $ where $\rho \geq 0 $ is the
within-cluster and between-cluster correlation coefficient between
$Y_{ij}(1,a)$ and $Y_{ij}(0,a)$. We can also rewrite $\sigma^2_w$ and
$\sigma^2_b$ as
$
\sigma^2_w \ = \ (1-r) \sigma^2$ and  $\sigma^2_b \ = \ r \sigma^2$,
where $\sigma^2=\sigma^2_w+\sigma_b^2$ represents the total variance
of $Y_{ij}(z,a)$ and $r = \sigma_b^2/(\sigma_w^2+\sigma_b^2)$ is the
intracluster correlation coefficient with respect to $Y_{ij}(z,a)$.
Denote $D_0^\ast = \text{diag}(D_{01}^\ast,D_{02}^\ast,\ldots, D_{0m}^\ast)$ with
\begin{eqnarray*}
D_{0a}^\ast \ = \ \frac{1}{q_a}\begin{pmatrix}
r+\frac{(1-p_a)(1-r)}{\bar{n}p_a} & \rho \left( r- \frac{1-r}{\bar{n}}\right)\\
 \rho \left( r- \frac{1-r}{\bar{n}}\right)& r+\frac{p_a(1-r)}{\bar{n}(1-p_a)}
\end{pmatrix}
\end{eqnarray*}
for $a=1,\ldots,m$, where $p_a$ is the treated proportion under
treatment assignment mechanism $a$ and $\bar{n}$ is harmonic mean of
$n_j$ defined as $\bar{n} = J/\sum_{j=1}^J \frac{1}{n_j}$.  When
$n_j=n$ for all $j$, $\bar{n} = n$.  Thus, $D_0^\ast$ is a
$2m \times 2m$ block diagonal matrix with $D_{0a}^\ast$ being the
$a$-th block for $a=1,\ldots,m$.
%

%

We derive the sample size formula for the direct effect under
Assumption~\ref{asm::simplification}. To reduce the number of
parameters in the alternative hypothesis $H_1: \ADE=x$, we consider
the alternative hypothesis about the  direct effects
across $m$ treatment assignment mechanisms:
\begin{eqnarray}
\label{eqn::alternative-de}
 H_1^{\text{de}}: |\ADE(a)| \ = \ \mu \quad \text{ for all } a.
\end{eqnarray}
The following theorem gives the sample size formula for rejecting the
null hypothesis $H_0: \ADE=0$, with respect to the alternative
hypothesis in Eqn.~\eqref{eqn::alternative-de}.  
\begin{theorem}[Simplified Sample Size Formula for Direct Effects] 
\label{thm::samplesize-de}
Consider a statistical hypothesis test with level $\alpha$ where the
null hypothesis is $H_0^{\text{de}}: \ADE=0$ and the
alternative hypothesis is given in
Eqn.~\eqref{eqn::alternative-de}. We reject the null hypothesis
if $T_{\text{de}} > \chi^2_{1-\alpha}(m)$ where the test statistic
$T_{\text{de}}$ is defined in Eqn.~\eqref{eqn::de-teststat}.
Under Assumption~\ref{asm::simplification}, the number of clusters
required for this test to have the statistical power of $1-\beta$ is
given by,
\begin{eqnarray}
\label{eqn::samplesize-de2}
J \ \geq \ \frac{ s^2( \chi^2_{1-\alpha}(m), 1-\beta,m)\cdot \sigma^2 }{\mu^2 } \cdot \frac{1}{ \sum_{a=1}^m \left\{(1,-1)D_{0a}^\ast(1,-1)^\top\right\}^{-1}}.
\end{eqnarray}
Moreover, if $r\geq 1/(n+1)$, then the required number of clusters is
given by,
\begin{eqnarray}
\label{eqn::samplesize-de3}
J \ \geq \ \frac{ s^2( \chi^2_{1-\alpha}(m), 1-\beta,m)\cdot \sigma^2 }{\mu^2 } \cdot \frac{1}{ \sum_{a=1}^m \left\{(1,-1)D_{0a}(1,-1)^\top\right\}^{-1}},
\end{eqnarray}
where  $D_{0a}= q_a^{-1} \textnormal{diag}\left(r+(np_a)^{-1} (1-p_a)(1-r), r+\{n(1-p_a)\}^{-1}p_a(1-r)\right) $.
\end{theorem}

To apply Eqn.~\eqref{eqn::samplesize-de2}, one needs to
  specify $(p_a, q_a)$ based on the study design and
  $(\rho,r,\sigma^2,\bar{n})$ based on prior information (e.g., pilot
  studies).  Because the sample size formula depends on the cluster
  sizes only through their harmonic mean, the formula can be applied
  regardless of whether the cluster sizes are given as fixed or
  random.  Since $\rho$ is
the correlation coefficient between potential outcomes under different
treatment conditions, it is an unidentifiable parameter. Therefore, we
provide a more conservative sample size formula in
Eqn.~\eqref{eqn::samplesize-de3} that does not involve $\rho$. The
condition $r\geq 1/(n+1)$ is easily satisfied so long as the cluster
size is moderate or large. Under this condition, if $J$ satisfies
Eqn.~\eqref{eqn::samplesize-de3}, then it also satisfies
Eqn.~\eqref{eqn::samplesize-de2}.

Next, we derive the sample size formula for the marginal direct effect
under Assumption~\ref{asm::simplification}. Because the marginal
direct effect is a scalar, we continue to use the alternative
hypothesis considered above, i.e., $H_1: \MDE=\mu$. The following
theorem gives the sample size formula.
\begin{theorem}[Simplified Sample Size Formula for Marginal Direct Effect] 
\label{thm::samplesize-mde}
Consider a statistical hypothesis test with level $\alpha$ where the
null hypothesis is $H_0^{\text{mde}}: \MDE =0$ and the alternative
hypothesis is $H_1^{\text{mde}}: \MDE=\mu$. We reject the null
hypothesis if $T_{\text{mde}} > \chi^2_{1-\alpha}(1)$ where
$T_{\text{mde}}$ is the test statistic defined in
Eqn.~\eqref{eqn::mde-teststat}. Under
Assumption~\ref{asm::simplification}, the number of clusters required
for the test to have the statistical power of $1-\beta$ is given by,
\begin{eqnarray}
\label{eqn::samplesize-mde2}
J \geq \frac{ s^2( \chi^2_{1-\alpha}(1), 1-\beta,1)\cdot \sigma^2 }{\mu^2 } \cdot \sum_{a=1}^mq_a^2\left\{(1,-1)D_{0a}^\ast(1,-1)^\top\right\}.
\end{eqnarray}
Moreover, if $r\geq 1/(n+1)$, then the number of clusters required is
given by, 
\begin{eqnarray}
\label{eqn::samplesize-mde3}
J \geq \frac{ s^2( \chi^2_{1-\alpha}(1), 1-\beta,1)\cdot \sigma^2 }{\mu^2 } \cdot \sum_{a=1}^mq_a^2\left\{(1,-1)D_{0a}(1,-1)^\top\right\}.
\end{eqnarray}
\end{theorem}
Similar to
Theorem~\ref{thm::samplesize-de}, the application of
Eqn.~\eqref{eqn::samplesize-mde2} requires the specification of
both $(p_a, q_a, n)$ and $(\rho,r,\sigma^2)$, while the more
conservative formula given in Eqn.~\eqref{eqn::samplesize-mde3}
does not depend on $\rho$.

Finally, we derive the sample size formula for the spillover effect
under Assumption~\ref{asm::simplification}. To reduce the number of
parameters in the alternative hypothesis $H_1: \ASE=x$, we consider the
following alternative hypothesis about the spillover effects
across different treatment conditions and treatment assignment
mechanisms,
\begin{eqnarray}
\label{eqn::alternative-se}
H_1^{\text{se}}: \max_{a\neq a'} |\ASE(z;a,a')| \ = \ \mu \quad\text{ for all } z.
\end{eqnarray}
The next theorem gives the sample size formula. 
\begin{theorem}[Simplified Sample Size Formula for Spillover Effects] 
\label{thm::samplesize-se}
Consider a statistical hypothesis test with level $\alpha$ where the
null hypothesis is $H_0^{\text{se}}: \ASE(z;a,a')=0$ for all $z$ and $a \neq a'$
and the alternative hypothesis given in
Eqn.~\eqref{eqn::alternative-se}. We reject the null hypothesis
if $T_{\text{se}} > \chi^2_{1-\alpha}(2(m-1))$ where the test
statistic $T_{\text{se}}$ is defined in
Eqn.~\eqref{eqn::se-teststat}. Under
Assumption~\ref{asm::simplification}, the number of clusters required
for the test to have the statistical power $1-\beta$ is given by,
\begin{eqnarray}
J \ \geq \ \frac{ s^2( \chi^2_{1-\alpha}(2(m-1)), 1-\beta,2(m-1))\cdot
 \sigma^2 }{\mu^2 \cdot \min_{s\in \mathcal{S}} s^\top \{C_3 D_0^\ast
 C_3^\top\}^{-1} s }, \label{eqn::samplesize-se}
\end{eqnarray}
where $\mathcal{S}$ is the set of $s =
(\ASE(0;1,2),\ASE(0;2,3),\ldots,
\ASE(0;m-1,m),\ASE(1;1,2),\ASE(1;2,3),\ldots,
\allowbreak\ASE(1;m-1,m))$ satisfying $\max_{a\neq a'}
|\ASE(z;a,a')| = 1$ for $z=0,1$. 
\end{theorem}
In Appendix~\ref{app:computation},
we show how to numerically compute the denominator of
Eqn.~\eqref{eqn::samplesize-se} using quadratic
programming. Unlike
Theorems~\ref{thm::samplesize-de}~and~\ref{thm::samplesize-mde}, we
cannot obtain a more conservative sample size formula by setting
$\rho$ to zero. Nonetheless, we use the following formula that does
not involve $\rho$ and evaluate its performance in our simulation
study given in Appendix~\ref{sec::simulation},
\begin{eqnarray}
\label{eqn::samplesize-se3}
J \ \geq \ \frac{ s^2( \chi^2_{1-\alpha}(2(m-1)), 1-\beta,2(m-1))\cdot \sigma^2 }{\mu^2 \cdot \min_{s\in \mathcal{S}} s^\top \{C_3 D_0 C_3^\top\}^{-1} s },
\end{eqnarray}
where $D_0 = \text{diag}(D_{01},D_{02},\ldots, D_{0m})$.
\vspace{-1cm}

\section{Empirical Analysis}
\label{sec::application}

\begin{figure}[t!]
\centering  
 \includegraphics[width=\textwidth]{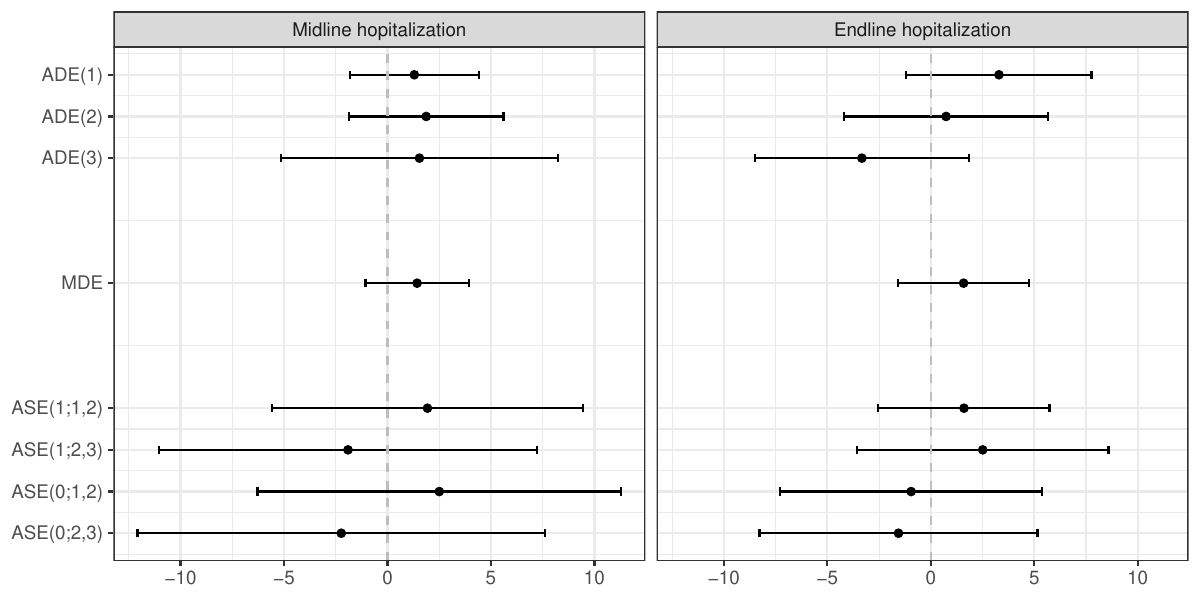}
 \caption{Estimated average direct, marginal direct, and spillover
   effects for the two outcomes of interest, midline hospitalization
   and endline hospitalization (percentage points). The top three lines
   are the average direct effects (ADE) under the three treatment
   assignment mechanisms; the
   middle line is the marginal direct effect (MDE); the bottom two
   lines are the average spillover effects (ASE) comparing the
   adjacent treatment assignment mechanisms under the
   treatment and control conditions. 95\% confidence intervals as well
   as point estimates are shown. }
\label{fig::APNE}
\end{figure}

In this section, we analyze the data from the randomized experiment of
the Indian national health insurance program described in
Section~\ref{sec::motivation}. We focus on two health outcomes:
midline and endline hospitalizations.  Figure~\ref{fig::APNE} shows
the estimated direct, marginal direct, and spillover effects for
midline and endline hospitalizations with their 95\% confidence
intervals.  For midline hospitalization (left panel), we find all
  of the estimated average direct effects to be positive under the
  three treatment assignment mechanisms but statistically
  insignificant.  Little heterogeneity in the direct effects means
  that the estimated marginal direct effect is similar to the three
  average direct effects. The spillover effects of treatment mechanism
  $1$ versus $2$ are estimated to be positive, while the spillover
  effects of treatment mechanism $2$ versus $3$ are estimated to be
  negative.  All of these spillover effects, however, are not
  statistically significant.

  For endline hospitalization (right panel), the estimated marginal
  direct effect is similar to that for midline
  hospitalization. However, heterogeneity exists across different
  treatment assignment mechanisms.  The estimated average direct
  effect is positive under treatment assignment mechanism $1$ but
  negative under treatment assignment mechanism $3$, with the
  difference between them being $6.6$ percentage points (95\% CI:
  $[-0.2,13.5]$). This may suggest that enrolling in the RSBY leads to
  a reduction in hospitalization in the long run, but only when the
  treatment proportion is not large. The spillover effects are
  positive under the treatment condition and negative under the
  control condition, but they are not distinguishable from zero.  


%
%
%
%
%

\begin{table}[t!]  
\caption{The required number of clusters for detecting the causal
 effects of certain sizes with the statistical power $0.8$ at the
 significance level $0.05$.}
\begin{center}
 \begin{tabular}{cccc}
 \hline
 &$|\ADE(a)| = 5\%$& $\MDE = 5\%$ & $ \max_{a\neq a'} |\ASE(z;a,a')| =5\%$\\ 
Midline hospitalization & $803$ & $585$ & $2230$ \\
Endline hospitalization & $400$ & $323$ & $857$ \\ \hline
\end{tabular}
\end{center}
\label{tab::APNE-samplesize}
\end{table}%

Next, we consider a hypothetical scenario, in which a researcher uses
this experiment as a pilot study for planning a future experiment.
The goal is to compute the sample size required for detecting certain
effect sizes at statistical power $0.8$ and significance level
$0.05$. For each outcome, we consider three null hypotheses:
$|\ADE(a)| = 5$ percentage points (pp.) for all $a$, $\MDE = 5$pp.,
and $ \max_{a\neq a'} |\ASE(z;a,a')| =5$pp. for all $z$. Note that the
total variance is $\sigma^2 = 0.175 $ for midline hospitalization and
$\sigma^2 = 0.180 $ for endline hospitalization; the intracluster
correlation coefficient is $r=0.42$ for midline hospitalization and
$r=0.11$ for endline hospitalization.

Table~\ref{tab::APNE-samplesize} presents the results.  We find that a
greater sample size is required for the midline hospitalization than
for the endline hospitalization. The reason is that the intracluster
correlation coefficient is much larger for the midline
hospitalization. In addition, a much greater sample size is required
for detecting the spillover effects than the direct effects. This is
because only a small proportion of the entire sample (i.e., 15\%) is
allocated to treatment assignment mechanism $2$. This leads to a
larger required overall sample size for detecting the spillover
effects.

\section{Concluding Remarks}
\label{sec::discussion}

In this paper, we introduced a general methodology for analyzing and
planning two-stage randomized experiments. Future research should
address several remaining methodological challenges. First, many
experiments suffer from attrition, which leads to missing outcome data
for some units. It is of interest to deal with such a complication in
the presence of spillover effects. Second, it is often believed that
spillover effects arise from interactions among a relatively small
number of units. How to explore this causal heterogeneity is an
important question to be addressed. Third, the standard two-stage
randomized design can be extended to sequential experimentation,
allowing researchers to examine how spillover effects evolve over
time. Finally, it is of interest to develop an optimal policy that
exploits spillover effects. The two-stage randomized design, or its
extensions, may be able to shed light on the construction of such
cost-effective policies.

\newpage

\pdfbookmark[1]{References}{References}
\spacingset{1.45}
\bibliographystyle{natbib}
\bibliography{interference,my,imai}

\newpage
\appendix

\setcounter{equation}{0}
\setcounter{figure}{0}
\setcounter{theorem}{0}
\setcounter{lemma}{0}
\setcounter{section}{0}
\renewcommand {\theequation} {S\arabic{equation}}
\renewcommand {\thefigure} {S\arabic{figure}}
\renewcommand {\thetheorem} {S\arabic{theorem}}
\renewcommand {\thelemma} {S\arabic{lemma}}
\renewcommand {\thesection} {S\arabic{section}}

\begin{center}
  \LARGE {\bf Supplementary Appendix}
\end{center}
Section~\ref{app::connection} establishes the equivalence relationship between the
regression-based inference and randomization-based inference.

Section~\ref{sec::comparison} compares the two-stage randomized design with the
 completely randomized and cluster randomized designs.

Section~\ref{app:proof} provides proofs of the theorems.

Section~\ref{app:computation} provides more computation details.

Section~\ref{sec::simulation} presents the simulation studies.

\section{Connections to linear regression}
\label{app::connection}

In this section, we establish direct connections between the proposed
estimators and the least squares estimators, which is popular among
applied researchers. \citet{basse2016analyzing} study the
relationships between the ordinary least squares and
randomization-based estimators for the direct and spillover effects
under a particular two-stage randomized experiment. Here, we extend
these previous results to a general setting with $m$ treatment
assignment mechanisms.

We consider the following linear model for the outcome,
\begin{eqnarray}
\label{eqn::regY}
Y_{ij}\ =\ \sum_{a=1}^m \left\{ \beta_{1a}Z_{ij} \bone(A_j=a) +\beta_{0a} (1-Z_{ij})\bone(A_j=a) \right\}+\epsilon_{ij},
\end{eqnarray}
where $\epsilon_{ij}$ is the error term. Unlike the two-step procedure
in \citet{basse2016analyzing}, we fit the weighted least squares
regression with the following inverse probability weights, 
\begin{eqnarray}
w_{ij} & = & \frac{1}{ J_{A_j}} \cdot \frac{1}{ n_{j Z_{ij}}}. \label{eq:ipw}
\end{eqnarray}
Let
$\widehat{\bm{\beta}}=(\widehat{\beta}_{11},\widehat{\beta}_{01},\ldots,\widehat{\beta}_{1m},\widehat{\beta}_{0m})^\top$
be the weighted least squares estimators of the coefficients in the
models of equation~\eqref{eqn::regY}, respectively. For the variance
estimator, we need additional notation. Let
$\bX_j=(X_{1j},\ldots, X_{n_j j})^\top$ be the design matrix of
cluster $j$ for the model given in~\eqref{eqn::regY} with
$X_{ij}=(Z_{ij}\bone(A_j=1),
(1-Z_{ij})\bone(A_j=1),\ldots,Z_{ij}\bone(A_j=m),
(1-Z_{ij})\bone(A_j=m))^\top$. Let
$\bX=(\bX_1^\top, \ldots, \bX_{J}^\top)^\top$ be the entire design
matrix, $\bW_j= \text{diag}(w_{1j},\ldots, w_{n_j j})$ be the weight
matrix for cluster $j$, and $\bW=\text{diag}(\bW_1,\ldots, \bW_J)$ be
the entire weight matrix. We use
$\hat\bepsilon_j =(\hat\epsilon_{1j},\ldots, \hat\epsilon_{n_jj})$ to
denote the residual vector for cluster $j$ obtained from the weighted
least squares fit of the model given in equation~\eqref{eqn::regY},
and
$\hat\bepsilon=(\hat\bepsilon_{1}^\top,\ldots,
\hat\bepsilon_{J}^\top)^\top$ to represent the residual vector for the
entire sample.

We consider the cluster-robust generalization of HC2 covariance matrix
\citep{bell:mcca:02},
\begin{eqnarray*}
\widehat{\var}^{\textnormal{cluster}}_{\textnormal{hc2}}(\widehat{\bm{\beta}})\ =\ (\bX^\top \bW \bX)^{-1} \left\{ \sum_j \bX_j^\top \bW_j (\bI_{n_j}-\bP_j)^{-1/2} \widehat{\bepsilon}_j \widehat{\bepsilon}_j^\top (\bI_{n_j}-\bP_j)^{-1/2} \bW_j \bX_j \right\} (\bX^\top \bW \bX)^{-1},
\end{eqnarray*}
where $\bI_{n_j}$ is the $n_j \times n_j$ identity matrix and $\bP_j$
is the following cluster leverage matrix,
\begin{eqnarray*}
 \bP_j \ = \ \bW_j^{1/2} \bX_j(\bX^\top \bW \bX)^{-1}
 \bX_j^\top \bW_j^{1/2}.
\end{eqnarray*}
The next theorem establishes the equivalence relationship between the
regression-based inference and randomization-based inference. 
\begin{theorem}[Equivalent Weighted Least Squares Estimators]
\label{thm::reg}
The weighted least squares estimators based on the model of
equation~\eqref{eqn::regY} are equivalent to the randomization-based
estimators of the average potential outcomes, i.e.,
$\widehat{\bm{\beta}} = \wY$. The cluster-robust generalization of
HC2 covariance matrix is equivalent to the randomization-based
covariance matrix estimator, i.e.,
$\widehat{\var}^{\textnormal{cluster}}_{\textnormal{hc2}}(\widehat{\bm{\beta}})=
\widehat D/J$.
\end{theorem}
Proof is given in Section~\ref{app::proof-thm9}.

\section{Theoretical Comparison of Three Randomized Experiments}
\label{sec::comparison}

Although the two-stage randomized design allows for the detection of
spillover effects, this may come at the cost of statistical efficiency
for detecting the average treatment effect if it turns out that
spillover effects do not exist. In this section, we conduct a
theoretical comparison of the two-stage randomized design with the
completely randomized design and cluster randomized design in the
absence of interference between units. The latter two are the most
popular experimental designs and are limiting designs of the two-stage
randomized designs. That is, we compute the relative efficiency loss
due to the use of the two-stage randomized design when there is no
spillover effect.

Formally, when there is no interference between units, we can write
$Y_{ij}(z,a)=Y_{ij}(z)$, $\oY_j(z,a)=\oY_j(z)$, and
$\oY(z,a)=\oY(z)$. As a result, both the direct and marginal direct
effects reduce to the standard average treatment effect. To unify the
notation in the three types of experiments, we define the unit-level
average treatment effect as, $\ATE_{ij}=Y_{ij}(1)-Y_{ij}(0)$, the
cluster-level average treatment effect as,
$\ATE_j= \sum_{i=1}^{n_j}\{Y_{ij}(1)-Y_{ij}(0)\}/n_j$, and the
population-level average treatment effect as
$\ATE= \sum_{i=1}^{J}\ATE_j/J$. As noted above, our comparison of
three designs assumes no interference between units. The reason for
this assumption is that the average treatment effect represents a
different causal quantity under the three designs in the presence of
interference, making the efficiency comparison across the designs less
meaningful \citep{karwa2018systematic}.

For simplicity, consider the case when the cluster size is equal,
i.e., $n_j=n$ for all $j$. Define the within-cluster variance of
$Y_{ij}(z)$ and $\ATE_{ij}$ as,
 \begin{eqnarray*}
\eta^2_w(z)&=& \frac{\sum_{j=1}^J\sum_{i=1}^{n}\{Y_{ij}(z)-\oY_j(z)\}^2}{nJ-1}, \quad \tau_w^2=\frac{\sum_{j=1}^J\sum_{i=1}^{n}\{\ATE_{ij}-\ATE_j\}^2}{nJ-1},
\end{eqnarray*}
the between-cluster variance of $Y_{ij}(z)$ and $\ATE_{ij}$ as,
\begin{eqnarray*}
 \eta^2_b(z)= \frac{\sum_{j=1}^{J}\{\oY_j(z)-\oY(z)\}^2}{J-1},\quad \tau_b^2=\frac{\sum_{j=1}^J\{\ATE_j-\ATE\}^2}{J-1},
\end{eqnarray*}
and the total variance of $Y_{ij}(z)$ and $\ATE_{ij}$ as,
\begin{eqnarray*}
\eta^2(z)&=& \frac{\sum_{j=1}^J\sum_{i=1}^{n}\{Y_{ij}(z)-\oY(z)\}^2}{nJ-1}, \quad \tau^2=\frac{\sum_{j=1}^J\sum_{i=1}^{n}\{\ATE_{ij}-\ATE\}^2}{nJ-1}.
\end{eqnarray*}
We can connect these variances by defining the intracluster correlation
coefficient with respect to $Y_{ij}(z)$ in cluster $j$ under treatment
condition $z$ as,
\begin{eqnarray*}
r_j(z)=\frac{\sum_{i\neq i'}^n (Y_{ij}(z)- \oY(z))(Y_{i'j}(z)- \oY(z))}{(n-1)\cdot \sum_{i=1}^n (Y_{ij}(z)- \oY(z))^2}.
\end{eqnarray*}
and the intracluster correlation coefficient with respect to $\ATE_{ij}$ in cluster $j$ as,
\begin{eqnarray*}
r'_j=\frac{\sum_{i\neq i'}^n (\ATE_{ij}- \ATE)(\ATE_{i'j}- \ATE)}{(n-1)\cdot \sum_{i=1}^n (\ATE_{ij}- \ATE)^2}.
\end{eqnarray*}

To further facilitate our theoretical comparison, we make additional
approximation assumptions. First, the intracluster correlation
coefficients are approximately the same with respect to $Y_{ij}(z)$
and $\ATE_{ij}$ across clusters and treatment conditions, i.e.,
$r_j(z) \approx r'_j \approx r$. Second, the cluster size is
relatively small compared to the number of clusters
$nJ-1 \approx nJ \approx n(J-1)$. These approximations help simplify
the expressions of the variances as
 \begin{eqnarray}
 \label{eqn::compare-approx}
\nonumber \eta_w^2(z) &\approx& \frac{(n-1)(1-r)}{n}\cdot \eta^2(z),\quad \tau_w^2\approx \frac{(n-1)(1-r)}{n}\cdot \tau^2,\\
 \eta_b^2(z) &\approx& \frac{1+(n-1)r}{n}\cdot \eta^2(z), \quad \tau_b^2\approx \frac{1+(n-1)r}{n} \cdot \tau^2.
\end{eqnarray}

%

We consider three randomized experiments in the population with $nJ$
units. Under the two-stage randomized design, the treatment is
randomized according to Assumptions~\ref{asm::two-stage}. Under the
completely randomized design, the treatment is randomized across
units,
\begin{eqnarray*}
 \Pr(\bZ= \bz) \ = \ \frac{1}{\binom{nJ}{\sum_{a=1}^m J_an p_a}},
\end{eqnarray*}
for all $\bz$ such that $\sum_{i,j} z_{ij} = \sum_{a=1}^m J_an p_a$.
Finally, under the clustered randomized design, the treatment is
randomized across clusters, where all the units in each cluster is
assigned to the same treatment condition, i.e.,
\begin{eqnarray*}
 \Pr(\bA= \ba) \ = \ \frac{1}{\binom{J}{\sum_{a=1}^m J_a p_a}},
\end{eqnarray*}
for all $\bz$ such that $\sum_{j=1}^J a_j =\sum_{a=1}^m J_a p_a$.
Note that under this setting, the number of treated units will be the
same in the three types of randomized experiments.

We consider the difference in means estimator for estimating $\ATE$,
\begin{eqnarray}
\widehat{\ATE}\ = \ \frac{1}{J} \sum_{j=1}^J \left\{ \frac{ \sum_{i=1}^n
 Y_{ij}Z_{ij}}{n_{j1}} - \frac{ \sum_{i=1}^n
 Y_{ij}(1-Z_{ij})}{n_{j0}}\right\}. \label{eqn::ATEestimator}
\end{eqnarray}
The following theorem gives the variances of this estimator under the
three experimental designs. 

%
%
\begin{theorem}[Comparison of Three Experimental Designs]
 \label{thm::comparison}   Under the approximation
 assumptions of equation~\eqref{eqn::compare-approx}, the variance of
 the average treatment effect estimator $\widehat{\ATE}$ given in
 equation~\eqref{eqn::ATEestimator} under the two-stage randomized
 design is
\begin{eqnarray}
\label{eqn::var:2stage-nointer}
\frac{1-r}{J^2} \sum_{a=1}^m \frac{J_a }{np_a} \cdot \eta^2(1) + \frac{1-r}{J^2} \sum_{a=1}^m \frac{J_a }{n(1-p_a)} \cdot \eta^2(0)- \frac{1-r}{nJ} \cdot \tau^2,
%
\end{eqnarray}
the variance of $\widehat \ATE$ under the completely randomized design is 
\begin{eqnarray}
\label{eqn::var:com-nointer}
&&\frac{1}{\sum_{a=1}^m J_an p_a} \cdot \eta^2(1)+\frac{1}{\sum_{a=1}^m J_an (1-p_a)} \cdot \eta^2(0)-\frac{1}{nJ} \cdot \tau^2,
%
\end{eqnarray}
the variance of $\widehat \ATE$ under the cluster randomized design is 
{\small
\begin{eqnarray}
\label{eqn::var:cluster-nointer}
\frac{1+(n-1)r}{\sum_{a=1}^m J_anp_a} \cdot \eta^2(1)+\frac{1+(n-1)r}{\sum_{a=1}^m J_an(1-p_a)} \cdot \eta^2(0)- \frac{1+(n-1)r}{nJ} \cdot \tau^2.
%
\end{eqnarray}
}
\end{theorem}
Proof is given in Appendix~\ref{app:comparison}. 
From Theorem~\ref{thm::comparison}, the ratio of the
coefficients of $\eta^2(1)$ in
equations~\eqref{eqn::var:2stage-nointer}
and~\eqref{eqn::var:com-nointer} is
\begin{eqnarray}
\label{eqn::ratio1}
 (1-r) \cdot \sum_{a=1}^m q_a p_a \cdot \sum_{a=1}^m \frac{q_a }{p_a},
\end{eqnarray}
whereas the ratio of the coefficients of $\eta^2(1)$ in
equations~\eqref{eqn::var:2stage-nointer}
and~\eqref{eqn::var:cluster-nointer} is
\begin{eqnarray}
\label{eqn::ratio2}
 \frac{1-r}{1+(n-1) r} \cdot \sum_{a=1}^m q_a p_a \cdot \sum_{a=1}^m \frac{q_a }{p_a}.
\end{eqnarray}
The ratios of the coefficients of other parameters take similar
forms. Thus, our discussion focuses on equations~\eqref{eqn::ratio1}
and~\eqref{eqn::ratio2}.

Equation~\eqref{eqn::ratio1} implies that the relative efficiency of
the two-stage randomized design over the completely randomized design
depends on the intracluster correlation coefficient, and the
assignment probabilities at the first and the second stage of
randomization. Due to the Cauchy--Schwarz inequality,
equation~\eqref{eqn::ratio1} is greater than or equal to $1-r$. The
value of this quantity increases as the heterogeneity between $p_a$
increases. Therefore, as the difference in treated proportions between
clusters becomes large, the two-stage randomized design becomes less
efficient for estimating the average treatment effect. On the other
hand, the ability to detect spillover effects relies on the
heterogeneity of $p_a$. This implies that there is a tradeoff between
the efficiency of estimating the average treatment effects and the
ability to detect spillover effects. This finding is consistent with
that of \citet{bair:etal:18}.

In addition, when the treated proportion is identical across clusters,
$p_a=p_{a'}$ for any $a, a'$, the two-stage randomized design becomes
stratified randomized design. In this case,
equation~\eqref{eqn::ratio1} equals $1-r$, which is less than
$1$. This is consistent with the classic result that the stratified
randomized design improves efficiency over the completely randomized
design.

Lastly, equation~\eqref{eqn::ratio2} implies that the relative
efficiency of the two-stage randomized design with respect to the
clustered randomized design depends additionally on the cluster
size. As the cluster size increases, the two-stage randomized design
becomes more efficient than the clustered randomized design. When
cluster size is large, the two-stage randomized design may be
preferable because it allows for the detection of spillover effects
while maintaining efficiency in estimating the average treatment
effect.

\section{Proofs of the Theorems}
\label{app:proof}
 We can write
\begin{eqnarray*}
\widehat{Y}(z,a) \ =\  \frac{1}{J_a}\sum_{j=1}^{J}\widehat{Y}_j(z)\bone(A_{j}=a) \ =\ \mu(z,a) + \sum_{j=1}^J \delta_j(z,a),
\end{eqnarray*}
where 
\begin{eqnarray*}
\mu(z,a) &=&  \frac{1}{J_a}\sum_{j=1}^{J}\oY_j(z,a)\bone(A_{j}=a),\\
\delta_j(z,a) & =&  \frac{1}{J_a} \left\{\widehat{Y}_j(z) - \oY_j(z,a) \right\}\bone(A_{j}=a).
\end{eqnarray*}
Let $\mu = (\mu(1,1),\mu(0,1),\ldots,\mu(1,m),\mu(0,m))^\top$ and $\delta_j = (\delta_j(1,1),\delta_j(0,1),\ldots,\delta_j(1,m),\delta_j(0,m))^\top$ be the vectorization of $\mu(z,a)$ and $\delta_j(z,a)$, respectively, and $\delta = \sum_{j=1}^J \delta_j$. We can write 
\begin{eqnarray*}
 \widehat{Y} \ = \  \mu + \delta.
\end{eqnarray*}
Let $\mathcal{F}_{0}=\sigma(A_1,\ldots,A_J)$ be the $\sigma$-algebra generated by $\{A_j: j =1,\ldots,J\}$.
Conditioning on $\mathcal{F}_{0}$, $\{\delta_j: j =1,\ldots,J \}$ are jointly independent. Therefore, we have $\E(\delta \mid \mathcal{F}_{0}) =\E(\delta_j \mid \mathcal{F}_{0})=0 $. From the law of total expectation,
\begin{eqnarray}
\nonumber
\E(\delta ) &=& \E(\delta_j ) \ =\  0,\\
\nonumber  \cov(\delta) &=& \sum_{j=1}^J \cov(\delta_j) \ = \ \sum_{j=1}^J \E \cov(\delta_j\mid  \mathcal{F}_{0}),\\
\label{eqn::cov-decomposition}
  \cov(\mu,\delta)&=& \E\left\{   \cov(\mu,\delta\mid  \mathcal{F}_{0})\right\}+ \cov\{\E(\mu\mid  \mathcal{F}_{0}),\E(\delta\mid  \mathcal{F}_{0})\}\ = \ 0.  
\end{eqnarray}



\subsection{Proof of Theorem~\ref{thm::var-general}}
\label{sec::covofwY}
We need the following lemma for the proof.

\begin{lemma}[\citet{li2017general}, Theorems 3 and 5]
\label{lem::CRT-CLT}
In a completely randomized experiment with $N$ units and $Q$ treatment groups of sizes $N_q$ $(q=1,\ldots,Q)$, 
let $Z_i$ be the treatment indicator and $Y_i$ be the observed outcome for unit $i$.
Let $Y_i(q)$ be the length-$L$ vector potential outcome of $i$ under treatment $q$, and $S_{qq'} = (N-1)^{-1}\sum_{i=1}^N\{Y_i(q)-\oY(q)\}\{Y_i(q')-\oY(q')\}$ be the finite-population covariances for $q,q'=1,\ldots,Q$. Let $\tau = \sum_{q=1}^Q G_q \oY(q)$ be the population average causal effect of interest, and $\widehat{\tau} =  \sum_{q=1}^WG_q \wY(q)$ be the moment estimator with  $\wY(q) = N_q^{-1} \sum_{i=1}^N Y_i \bone(Z_i=q)$. We have (a)
\begin{eqnarray*}
\cov(\widehat{\tau})\ = \ \sum_{q=1}^Q N_q^{-1}G_qS_{qq}G_q^\top-N^{-1} S^2_{\tau},
\end{eqnarray*}
where $S^2_\tau$ is the finite-population covariance of $\tau_i=\sum_{q=1}^WG_q Y_i(q)$; (b) suppose the following conditions hold for  $q,q'=1,\ldots,Q$ as $N$ goes to infinity:
\begin{enumerate}[(a)]
\item $S_{qq'}$ has a finite limit;
\item $N_q/N$ has a finite limit in $(0,1)$;
\item $\max_{i} |Y_i(q)-\oY(q)|^2/N = o(1)$. Then, we have 
\begin{eqnarray*}
\sqrt{N}(\widehat{\tau}-\tau) \stackrel{d}{\rightarrow}  N(0,V),
\end{eqnarray*}
where $V$ denotes the limiting value of $N\cov(\widehat{\tau})$.
\end{enumerate}
\end{lemma}

\medskip

We then prove  Theorem~\ref{thm::var-general}.
For simplicity, we consider the case with $m=2$.
From~\eqref{eqn::cov-decomposition}, we have  $\cov(\wY) =  \cov(\mu)+\cov(\delta)$. We then derive the analytic forms of $\cov(\mu)$ and $\cov(\delta)$. 

For $\cov(\mu)$, define $B_j(a) = (\oY_j(1,a),\oY_j(0,a))^\top$ as the vector potential outcome of cluster $j$ under $A_j=a$ with means $\overline{B}(a) = (\oY(1,a),\oY(0,a))^\top$ and covariances
\begin{eqnarray*}
S_b(a) &= & (J-1)^{-1} \sum_{j=1}^J\{B_j(a) -\overline{B}_j(a)  \}\{W_j(a) -\overline{B}_j(a)  \}^\top\\
&=& \begin{pmatrix}
\sigma^2_b(1,1;a,a) & \sigma^2_b(1,0;a,a)\\
\sigma^2_b(1,0;a,a) & \sigma^2_b(0,0;a,a)
\end{pmatrix}.
\end{eqnarray*}
We can then write  $\mu$ as  $(I_2,0_{2 \times 2})^\top \overline{B}(1)+(0_{2 \times 2},I_2)^\top \overline{B}(2)$. From Lemma~\ref{lem::CRT-CLT},  we have 
\begin{eqnarray*}
\cov(\mu) &=& J_1^{-1}  (I_2,0_{2 \times 2})^\top S_b(1) (I_2,0_{2 \times 2})+J_2^{-1}  (0_{2 \times 2},I_2)^\top S_b(2) (0_{2 \times 2},I_2) - J^{-1} S_b\\
&=& J^{-1} (H \circ S_b).
\end{eqnarray*}

For $\cov(\delta)$, theory of simple random sampling implies,
\begin{eqnarray*}
\var \left\{\widehat{Y}_j(z,a) \mid A_j=a\right \} &=&  \frac{1}{n_{jz}} \left(1- \frac{n_{jz}}{n_j} \right) \sigma^2_j(z,a),\\
\cov \left\{\widehat{Y}_j(1,a), \widehat{Y}_j(0,a) \mid A_j=a\right \} &=& -\frac{1}{n_j} \sigma^2_j(1,0;a).
\end{eqnarray*}
We can write
\begin{eqnarray*}
\cov \left\{\widehat{Y}_j(z,a), \widehat{Y}_j(z',a) \mid A_j=a\right \}&=& n_j^{-1} \{n_j/n_{jz} \bone(z=z')-1 \}  \sigma^2_j(z,z';a).
\end{eqnarray*}
Therefore, we have
\begin{eqnarray}
\label{eqn::covdelta-A}
\cov(\delta_j\mid A_j =a )\ = \ J_a^{-2}q_a n_j^{-1} \{ H_j(a) \circ S_w\} \ =\ q_a^{-1}J^{-2}n_j^{-1} \{ H_j(a) \circ S_j\},
\end{eqnarray}
where 
\begin{eqnarray*}
H_j(1) &=& \text{diag}(q_1^{-1},0 ) \otimes \left\{ \text{diag}(n_j/n_{j1},n_j/n_{j0} )-1_{2\times 2}  \right\},\\
H_j(2)&=&\text{diag}(0,q_2^{-1} ) \otimes \left\{ \text{diag}(n_j/n_{j1},n_j/n_{j0} )-1_{2\times 2}  \right\}.
\end{eqnarray*}
 Because $H_j =H_j(1)+H_j(2) $, we can obtain
\begin{eqnarray*}
\cov(\delta_w) \ = \ \E\{\cov(\delta_j\mid A_j=a)\}\ = \ J^{-2}n_j^{-1} \{ H_j \circ S_j\}.
\end{eqnarray*}
\QEDB

\subsection{Proof of Theorem~\ref{thm::varest}}
\label{app::conservative-cov}

Recall that $\widehat D$ be a $2m$ by $2m$ block diagonal matrix
with the $a$-th matrix on the diagonal 
$$
\widehat D_a = \frac{J}{J_a}\begin{pmatrix}
 \widehat \sigma^2_b(1,a) &  \widehat  \sigma^2_b(1,0;a) \\
 \widehat \sigma^2_b(1,0;a) &  \widehat \sigma^2_b(0,a)
\end{pmatrix}.
$$ 
We calculate the expectation of each term in $\widehat D_a$. 
We have 
\begin{eqnarray*}
&&\E \{\widehat{\sigma}_{b}^2(z,a)\}\\
& = & \frac{1}{J_a-1} \E \left\{ \sum_{j=1}^J \widehat{Y}^2_j(z) I(A_j=a)- J_a \widehat{Y}(z)^2 \right\}\\
&=& \frac{1}{J_a-1} \E \left( \sum_{j=1}^J\left [ \var \left\{ \widehat{Y}_j(z,a) \mid A_j=a\right\}+ \oY_j(z,a)^2 \right] I(A_j=a)\right)- \frac{J_a}{J_a-1} [ \var \{\widehat{Y}(z,a)\} + \oY(z,a)^2]\\
&=&  \frac{J_a}{J(J_a-1)} \sum_{j=1}^J \var \{ \widehat{Y}_j(z) \mid A_j=a\}+\frac{J_a}{J(J_a-1)} \sum_{j=1}^J \oY_j(z,a)^2-\frac{J_a}{J_a-1} [ \var \{\widehat{Y}(z,a)\} + \oY(z,a)^2]\\
&=&  \frac{J_a}{J(J_a-1)} \sum_{j=1}^J \var \left\{ \widehat{Y}_j(z) \mid A_j=a\right\} + \frac{J_a(J-1)}{(J_a-1)J} \sigma_{b}^2(z,a)- \frac{J_a}{J_a-1} \var \{\widehat{Y}(z,a)\} \\
&=&  \frac{J_a}{J(J_a-1)} \sum_{j=1}^J \var\left \{ \widehat{Y}_j(z) \mid A_j=a\right\} + \frac{J_a(J-1)}{(J_a-1)J} \sigma_{b}^2(z,a)\\
&&- \frac{J_a}{J_a-1} \left[ \left(1- \frac{J_a}{J} \right) \frac{\sigma_{b}^2(z,a) }{J_a} +\frac{1}{J_a J} \sum_{j=1}^J \var \left\{ \widehat{Y}_j(z,1) \mid A_j=a \right \}\right] \\
&=&\sigma_{b}^2(z,a) + \frac{1}{J} \sum_{j=1}^J \var \left\{ \widehat{Y}_j(z,a) \mid A_j=a\right\}\\
&=& \sigma_{b}^2(z,a) + \frac{1}{J} \sum_{j=1}^J \frac{1}{n_{jz}} \left(1- \frac{n_{jz}}{n_j} \right) \sigma^2_j(z,a).
\end{eqnarray*}
Similarly, we obtain
\begin{eqnarray*}
\E \left\{\widehat{\sigma}_{b}^2(1,0;a)\right\}
&=&\sigma_{b}^2(1,0;a) + \frac{1}{J} \sum_{j=1}^J \cov \left\{ \widehat{Y}_j(1,a),\widehat{Y}_j(0,a) \mid A_j=a\right\}\\
&=&\sigma_{b}^2(1,0;a) - \frac{1}{J} \sum_{j=1}^J\frac{\sigma^2_j(1,0;a)}{n_j} .
\end{eqnarray*}

Finally, we prove that $\widehat D$ is a conservative estimator for
$D$. Denote $R = \E(\widehat D)-D$ with the $(k,l)$-th element
$r_{kl}$. We have
\begin{eqnarray*}
r_{2a-1,2a-1} &=& \sigma^2_b(1,a),\quad r_{2a,2a} = \sigma^2_b(0,a),\quad
r_{2a-1,2a} = \sigma^2_b(1,0;a)
\end{eqnarray*}
 for $a=1,\ldots,m$. For $a\neq a'$, we have 
 \begin{eqnarray*}
r_{2a-1,2a'-1} &=& \sigma^2_b(1;a,a'),\quad r_{2a,2a'-1} =\sigma^2_b(1,0;a,a').
\end{eqnarray*}
Therefore, for any vector $c = (c_1,\ldots, c_{2m})$, $cRc^\top$ is the between-cluster variance of 
$\sum_{a=1}^m c_{2a-1} Y_{ij}(1,a)+\sum_{a=1}^m c_{2a}Y_{ij}(0,a)$. As a result, $\widehat D$ is a conservative estimator for $D$ and is unbiased for $D$ if $\oY_j(z,a)$ is constant across clusters.
 \QEDB


\subsection{Proof of Theorem~\ref{thm::CLT}}
\label{app::CLT}

\subsubsection{Lemmas}
Let $*$ denote convolution.
We need the following lemmas for the proof.

\begin{lemma}[\citet{ohlsson1989asymptotic}, Theorem A.1]
\label{lem::2stageCLT}
For $j=1,\ldots,J$, let $\{ \xi_{J,j}: j=1,\ldots,J \}$ be a martingale difference sequence relative to the filtration $\{\mathcal{F}_{J,j}: j =0,1,\ldots,J\}$, and let $X_J$ be an $\mathcal{F}_{J,0}$-measurable random variable.  Denote $\xi_J=\sum_{j=1}^J\xi_{J,j} $. Suppose that the following conditions hold as $J$ goes to infinity:
\begin{enumerate}[(a)]
\item $\sum_{j=1}^J \E(\xi_{J,j}^4) = o(1)$.
\item For some sequence of non-negative real numbers $\{\beta_J: J=1,2,\ldots\}$ with $\text{sup}_J \beta_J < \infty$, we have $\E \left[\left\{\sum_{j=1}^J\E(\xi_{J,j}^2\mid \mathcal{F}_{J,j-1}) - \beta_J^2  \right\}^2\right] = o(1)$.
\item For some probability distribution $\mathcal{L}_0$, $\mathcal{L}(X_J)*N(0,\beta_J^2) \stackrel{d}{\rightarrow}  \mathcal{L}_0 $.
\end{enumerate}
Then, $\mathcal{L}(X_J+\xi_J) \stackrel{d}{\rightarrow}  \mathcal{L}_0$ as $J$ goes to infinity.
\end{lemma}

\begin{lemma}
\label{lem::1}
Suppose that Assumptions~\ref{asm::two-stage},~\ref{asm::nointer},~\ref{asm::stratified},~and Condition~\ref{con::regularity} hold. Then
\begin{eqnarray*}
J^2 \sum_{j=1}^J \E(||\delta_j||_2^4 \mid A_j=a)\ =\ o(1),\quad  J^2 \sum_{j=1}^J \E(||\delta_j||_2^4 )\ =\ o(1)
\end{eqnarray*}
\end{lemma}
\noindent {\it Proof.} It suffices to verify the first equality. Note that for $j$ with $A_j=a$ ($a=1,2$) 
$$
||\delta_j||_2^2\ =\ \frac{1}{J_a^2} \left[ \left\{\widehat{Y}_j(1) - \oY_j(1,a) \right\}^2+\left\{\widehat{Y}_j(0) - \oY_j(0,a) \right\}^2 \right].
$$
From the Cauchy--Schwarz inequality, we have 
\begin{eqnarray*}
||\delta_j||_2^4 \leq  \frac{2}{J_a^4} \left[ \left\{\widehat{Y}_j(1) - \oY_j(1,a) \right\}^4+\left\{\widehat{Y}_j(0) - \oY_j(0,a) \right\}^4 \right].
\end{eqnarray*}
Because $ \left\{\widehat{Y}_j(z) - \oY_j(z,a) \right\}^4 \leq 8 \wY^4_j(z)+8 \oY^4_j(z)$,
\begin{eqnarray*}
&&\sum_{j=1}^J\E \left[  \left\{\widehat{Y}_j(z) - \oY_j(z,a) \right\}^4 \mid A_j=a\right]\\
&=& 8  \sum_{j=1}^J\E \left\{ \widehat{Y}_j^4(z)  \mid A_j=a\right\}+ 8  \sum_{j=1}^J  \oY^4_j(z,a)\\
&=&8  \sum_{j=1}^J\E \left\{ \widehat{Y}_j^4(z)  \mid A_j=a\right\} + o(J^2).
\end{eqnarray*}
From the power-mean inequality, for $A_j=a$,
\begin{eqnarray}
\label{eqn::ineq-Y4}
 \widehat{Y}_j^4(z)\  \leq \ \frac{1}{n_{jz}} \sum_{j=1}^J Y^4_{ij}(z,a) \cdot \bone(Z_{ij}=z)\ \leq\   \frac{1}{n_{jz}} \sum_{j=1}^J Y^4_{ij}(z,a)\ \leq \ \epsilon^{-1} \overline{Y_j(z,a)}\ =\ o(J^2).
\end{eqnarray}
Therefore,
\begin{eqnarray*}
\sum_{j=1}^J\E \left[  \left\{\widehat{Y}_j(z) - \oY_j(z,a) \right\}^4 \mid A_j=a\right]\ = \ o(J^2).
\end{eqnarray*}
As a result,
\begin{eqnarray*}
J^2 \sum_{j=1}^J \E(||\delta_j||_2^4 \mid A_j=a) &\leq& \frac{2}{q_a^2 J^2} \sum_{z=0,1} \sum_{j=1}^J\E \left[  \left\{\widehat{Y}_j(z) - \oY_j(z,a) \right\}^4 \mid A_j=a\right]\\
&=& o(1).
\end{eqnarray*}
\QEDB

\subsubsection{Proof of the asymptotic normality}
For simplicity, we focus on the case with $m=2$.
We only need to show that for any unit vector $\eta$ with length $4$,
\begin{eqnarray*}
\eta^\top \sqrt{J}(\widehat{Y}-\oY) = \eta^\top \sqrt{J}(\mu-\oY+\delta)  \stackrel{d}{\rightarrow}  N(0, \eta^\top D \eta).
\end{eqnarray*}
Let  $X_J = \eta^\top \sqrt{J}(\mu - \oY)$ and $\xi_{J,j} = \eta^\top \sqrt{J}\delta_j$. 
It suffices to verify the conditions in Lemma~\ref{lem::2stageCLT}. We will suppress $J$ in the subscripts when no confusion arises.

First, $\mathcal{F}_{J,0}$ contains the information from the first stage randomization, and $\mathcal{F}_{J,j}$ contains the information from the first stage randomization plus the second stage randomization in the first $j$ clusters. Therefore, $\{\mathcal{F}_{J,j}: j = 0,\ldots,J\}$ is a filtration.

For Lemma~\ref{lem::2stageCLT} condition (a), from the Cauchy--Schwarz inequality, we have 
\begin{eqnarray*}
\xi_{J,j}^4\ =\ J^2 (\eta^\top \delta_j)^4 \leq J^2 || \eta||_2^4 \cdot ||\delta_j||_2^4\ = \ J^2 \cdot ||\delta_j||_2^4.
\end{eqnarray*}
From Lemma~\ref{lem::1}, we have 
\begin{eqnarray*}
\sum_{j=1}^J \E(\xi_{J,j}^4) \ \leq\  J^2 \sum_{j=1}^J \E(||\delta_j||_2^4) \ = \ o(1).
\end{eqnarray*}

For Lemma~\ref{lem::2stageCLT} condition (b), we have from Theorem~\ref{thm::var-general},
$$
\beta_J^2 \ = \ \var(\xi_J)\ = \ J^{-1} \eta^\top \left\{ \sum_{j=1}^J  n_j^{-1} \{ H_j \circ S_j\}\right\} \eta.
$$
Because $\E\{\xi^2_{J,j}\mid \mathcal{F}_{J,j-1}\}=\E\{\xi^2_{J,j}\mid \mathcal{F}_{J,0}\} =\var\{\xi_{J,j}\mid \mathcal{F}_{J,0}\} $ and $\E(\xi_{J,j}\mid \mathcal{F}_{J,0})=0$, we have
\begin{eqnarray*}
\sum_{j=1}^J\E\{\xi^2_{J,j}\mid \mathcal{F}_{J,j-1}\}\ = \ \sum_{j=1}^J\var\{\xi_{J,j}\mid \mathcal{F}_{J,0}\} \ = \ \var(\xi_J \mid \mathcal{F}_{J,0})
\end{eqnarray*}
and $\beta_J^2 = \E \{ \var(\xi_J \mid \mathcal{F}_{J,0}) \}$. Therefore,
$$
\E \left[\left\{\sum_{j=1}^J\E(\xi_{J,j}^2\mid \mathcal{F}_{J,j-1}) - \beta_J^2  \right\}^2\right]\ =\ \var\left\{ \var(\xi_J \mid \mathcal{F}_{J,0})\right\}.
$$
Therefore, we only need to verify that $ \var\left\{ \var(\xi_J \mid \mathcal{F}_{J,0})\right\}=o(1)$.
Denote 
\begin{eqnarray*}
\zeta_j(a) \ = \ \E(\xi^2_{J,j}\mid A_j=a) \ = \ J\eta^\top \cov(\delta_j \mid A_j=a) \eta
\end{eqnarray*}
with mean $\overline{\zeta}(a) = J^{-1} \sum_{j=1}^J \zeta_j(a)$, variance $S_{\zeta(a)}=(J-1)^{-1} \sum_{j=1}^J \{\zeta_j(a)-\overline{\zeta}(a)\}^2 $, and sample mean $\widehat{\zeta}(a) = J_a^{-1}\sum_{j=1}^J \zeta_j(a) \bone(A_j=a) $ for $a=1,2$. We have $\var(\xi_J \mid \mathcal{F}_{J,0}) = J_1\widehat{\zeta}(1)+J_2 \widehat{\zeta}(2)   $. Theory of simple random sampling implies $\var\{\widehat{\zeta}(a)\} = (1-q_a)q_a^{-1}S_{\zeta(a)}$. Therefore, we have 
\begin{eqnarray*}
\var\left\{ \var(\xi_J \mid \mathcal{F}_{J,0})\right\} &=& \var\left\{ J_1\widehat{\zeta}(1)+J_2 \widehat{\zeta}(2)  \right\}\\
&\leq & 2\var\left\{ J_1\widehat{\zeta}(1)\right\}+ 2\var\left\{ J_1\widehat{\zeta}(1)\right\}\\
&=&2Jq_1q_2(S_{\zeta(1)}+S_{\zeta(2)}).
\end{eqnarray*}
From the Cauchy--Schwarz inequality, 
\begin{eqnarray*}
\zeta^2_j(a)\ = \ \{\E(\xi_j^2\mid A_j=a)\}^2\ \leq \ \E(\xi_j^4\mid A_j=a)\ \leq \ J^2  \E\{ ||\delta_j||_2^4\mid A_j=a).
\end{eqnarray*}
Thus, from Lemma~\ref{lem::1}, we have 
\begin{eqnarray}
\label{eqn::zeta1}
(J-1)^{-1} \sum_{j=1}^J \{\zeta_j(a)\}^2 \ \leq \  (J-1)^{-1}  J^2 \sum_{j=1}^J  \E\{ ||\delta_j||_2^4\mid A_j=a) \ = \ o(J^{-1}).
\end{eqnarray}
From~\eqref{eqn::covdelta-A}, we have 
\begin{eqnarray}
\nonumber \overline{\zeta}(a)  &=&  \eta^\top \left\{\sum_{j=1}^J \cov(\delta_j \mid A_j=a) \right\}\eta\\
\nonumber &=&  \eta^\top \left\{q_a^{-1}J^{-2}\sum_{j=1}^J n_j^{-1} \{ H_j(a) \circ S_w\} \right\}\eta\\
\label{eqn::zeta2} &=& o(J^{-1}),
\end{eqnarray}
where the last equality follows from Condition~\ref{con::regularity}.
Combining~\eqref{eqn::zeta1}~and~\eqref{eqn::zeta2}, we have $S_{\zeta(a)} = o(J^{-1})$ for $a=1,2$, which leads to $\var\left\{ \var(\xi_J \mid \mathcal{F}_{J,0})\right\}=o(1)$.

We then consider Lemma~\ref{lem::2stageCLT} condition (c). From Lemma~\ref{lem::CRT-CLT}~and Theorem~\ref{thm::var-general}, we have 
$ \sqrt{J}(\mu - \oY) \stackrel{d}{\rightarrow}  N(0, H \circ S_b)$ under Condition~\ref{con::regularity}. Thus the convolution of $\mathcal{L}(X_J)$  with $N(0,  \eta^\top\left\{\sum_{j=1}^JJ^{-2}n_j^{-1} ( H_j \circ S_w) \right\}\eta )$ converges in distribution to $N(0,\eta^\top D \eta)$.
\QEDB

\subsection{Proof of Theorem~\ref{thm::test}}
\label{app::test}
We need the following two lemmas.
\begin{lemma}
\label{lem::consistency-Y}
Suppose that
 Assumptions~\ref{asm::two-stage},~\ref{asm::nointer},~\ref{asm::stratified},~and Condition~\ref{con::regularity} hold. Then $\widehat{D}-\E\{\widehat{D}\}=o(1)$ a.s.
\end{lemma}
\noindent {\it Proof of Lemma~\ref{lem::consistency-Y}.} Denote 
 \begin{eqnarray*}
\widehat{T}(z,z';a)\ =\ J_a^{-1} \sum_{j=1}^J  \wY_j(z) \wY_j(z') \bone(A_j=a).
\end{eqnarray*}
We first show that $\widehat{T}(z,z';a) - \E\left\{ \widehat{T}(z,z';a)\right\} = o(1)$.  It suffices to verify that $\cov\left\{ \widehat{T}(z,z';a) \right\} = o(1)$.
 Denote $U_j = \wY_j(z) \wY_j(z') \bone(A_j=a) $ and $\mu_j = \E(U_j\mid A_j =a )$. We can write
 \begin{eqnarray*}
\cov\left\{ \widehat{T}(z,z';a) \right\} \ = \ J_a^{-2} \left\{\sum_{j=1}^J \cov(X_j)  + \sum_{j\neq k} \cov(X_j,X_k) \right\}.
\end{eqnarray*}
By some algebra, we have
\begin{eqnarray*}
\E\{\cov(X_j \mid A_j) \}&=& q_a  \cov(X_j \mid A_j=a) \ = \ q_a \E(X_j^2\mid A_j=a) - q_a \mu_j^2,\\
\cov\{\E(X_j\mid A_j) \} & =& \E \left[ \{\E(X_j\mid A_j) \}^2\right] - \{\E(X_j)\}^2\\
&=& q_a\{\E(X_j \mid A_j =a) \}^2 - q_a^2\mu_j^2\\
&=& q_a \mu_j^2 -  q_a^2\mu_j^2\\
\E\{\E(X_j\mid A_j)\E(X_k\mid A_k)  \} &=& \Pr(A_k=A_j=a) \E(X_j \mid A_j =a)\E(X_k \mid A_k =a)\\
&=&q_a \frac{J_a-1}{J-1} \mu_j\mu_k,\\
\E(X_j)\E(X_k) &=& q_a^2 \mu_j\mu_k.
\end{eqnarray*}
Therefore,
\begin{eqnarray*}
\cov(X_j) &=& \E\{\cov(X_j \mid A_j) \}+\cov\{\E(X_j\mid A_j) \} \\
&=& q_a \E(X_j^2\mid A_j=a)- q_a^2\mu_j^2,\\
\cov(X_j,X_k) &=& \cov\{\E(X_j\mid A_j),\E(X_k\mid A_k)  \}  + \E\{\cov(X_j,X_k \mid A_j,A_k)\}\\
&=& \cov\{\E(X_j\mid A_j),\E(X_k\mid A_k)  \} \\
&=& \E\{\E(X_j\mid A_j)\E(X_k\mid A_k)  \}  - \E(X_j)\E(X_k)\\
&=& - (J-1)^{-1}q_a(1-q_a)\mu_j\mu_k.
\end{eqnarray*}
As a result, we have
\begin{eqnarray*}
&&J_a^2\cov\left\{ \widehat{T}(z,z';a) \right\}\\
& =& q_a \sum_{j=1}^J\E(X_j^2\mid A_j=a) - q_a^2  \sum_{j=1}^J\mu_j^2-\frac{q_a(1-q_a)}{J-1}\sum_{j\neq k} \mu_j\mu_k\\
& =& q_a \sum_{j=1}^J\E(X_j^2\mid A_j=a) - q_a^2  \sum_{j=1}^J\mu_j^2+\frac{q_a(1-q_a)}{J-1}\sum_{j=1}^J \mu_j^2 -\frac{q_a(1-q_a)}{J-1}\sum_{j, k} \mu_j\mu_k\\
&\leq &   q_a \sum_{j=1}^J\E(X_j^2\mid A_j=a) - \left\{ q_a^2-\frac{q_a(1-q_a)}{J-1}  \right\} \sum_{j=1}^J\mu_j^2.
\end{eqnarray*}
When $J$ goes to infinity,  we can obtain
\begin{eqnarray*}
J_a^2\cov\left\{ \widehat{T}(z,z';a) \right\} \ \leq \ q_a \sum_{j=1}^J\E(X_j^2\mid A_j=a) \ = \  q_a \sum_{j=1}^J\E\left\{\wY^2_j(z)\wY^2_j(z')\mid A_j=a\right\}.
\end{eqnarray*}
From  $\wY^2_j(z)\wY^2_j(z') \leq 2 \wY^4_j(z)+2\wY^4_j(z')$ and~\eqref{eqn::ineq-Y4}, we then have
\begin{eqnarray*}
J_a^2\cov\left\{ \widehat{T}(z,z';a) \right\} \ \leq \ 2q_a \sum_{j=1}^J\E\left\{\wY^4_j(z)+\wY^4_j(z')\mid A_j=a\right\}\ = \ o(1),
\end{eqnarray*}
where the last equality follows from a similar argument in the proof of Lemma~\ref{lem::1}. Therefore, we have $\widehat{T}(z,z';a) - \E\left\{ \widehat{T}(z,z';a)\right\} = o(1)$. 

We then show that  $\wY-\oY= o(1)$. From Theorem~\ref{thm::var-general}, we have 
\begin{eqnarray*}
\cov(\wY)\ =\ J^{-1} (H \circ S_b)+J^{-2} \sum_{j=1}^J n_j^{-1} \{ H_j \circ S_j\} \ = \ o(1),
\end{eqnarray*}
where the last equality follows from Condition~\ref{con::regularity}.

Finally, we prove that $\widehat{D}-\E(\widehat{D})= o(1)$ as $J$ goes to infinity.  The elements of $\widehat{D}$ are $J/J_a \widehat{\sigma}^2(z,z';a)$, which can be written as
\begin{eqnarray*}
q_a^{-1}  \left\{\widehat{T}(z,z';a) - \wY(z,a)\wY(z,a')\right\},
\end{eqnarray*}
where we ignore the difference between $J_a$ and $J_a-1$.
Therefore, $J/J_a \widehat{\sigma}^2(z,z';a)$ converges to
\begin{eqnarray*}
q_a^{-1}  \left[\E\left\{\widehat{T}(z,z';a)\right\} - \oY(z,a)\oY(z,a')\right],
\end{eqnarray*}
which is equal to  $J/J_a  \E\left\{\widehat{\sigma}^2(z,z';a)\right\}$.\QEDB

\medskip

 \begin{lemma}
 \label{lem::chisquare}
 (i) If $X \sim \text{N}_k(0,A)$, Then $X^\top B X \stackrel{d}{=}\sum_{j=1}^k\lambda_j(AB)\xi_j^2$, where  the $\lambda_j(AB)$'s are eigenvalues of $AB$, and $\xi_j \sim \chi^2(1)$ and are i.i.d. with each other.\\
 (ii) If $X_n \stackrel{d}{\rightarrow} \text{N}_k(0,A)$, and $B_n \stackrel{p}{\rightarrow} B $, then $X_n^\top B_n X_n \stackrel{d}{=}\sum_{j=1}^k\lambda_j(AB^{-1})\xi_j^2$. If $B-A$ is positive semidefinite, then $0\leq \lambda_j(AB^{-1})\leq 1$ for all $j$.
 \end{lemma}
 \noindent {\it Proof of Lemma~\ref{lem::chisquare}}. Lemma~\ref{lem::chisquare}(i) follows form linear algebra and Lemma~\ref{lem::chisquare}(ii) follows from Slutsky's Theorem. \QEDB

\medskip

We then prove  Theorem~\ref{thm::test}.
From Theorem~\ref{thm::CLT}~and~Lemma~\ref{lem::consistency-Y}, we know that
$\sqrt{J} (C\wY-x) \stackrel{d}{\rightarrow} \text{N}_{2m}(0,CDC^\top)$,
$C\widehat D C^\top \stackrel{p}{\rightarrow} C \E(\widehat D) C^\top
$. Because $C \E(\widehat D) C^\top -C D C^\top $ is positive
semi-definite, from Lemma~\ref{lem::chisquare}(ii), we have
$T \stackrel{d}{=}\sum_{j=1}^k\lambda_j\xi_j^2$, where $k$ is the rank
of $C$ and $0\leq \lambda_j\leq 1$ for all $j$. \QEDB

\subsection{Proof of Theorem~\ref{thm::samplesize-general}}
\label{app::general-samplesize}
%
To prove Theorem~\ref{thm::samplesize-general}, we need the following lemma.
\begin{lemma}
\label{lem::chi}
Suppose $(X_1,\ldots,X_k)$ follows a standard multivariate normal distribution.
If $0< a_j \leq a'_j$ for $j=1,\ldots,k$, then as $J$ goes to infinity,
\begin{eqnarray*}
 \Pr\left\{ \sum_{j=1}^k \left( a'_j X_j+ \sqrt{J} x_j \right)^2 \geq t \right\} \geq p
\end{eqnarray*}
implies
\begin{eqnarray*}
 \Pr\left\{ \sum_{j=1}^k \left( a_j X_j+ \sqrt{J} x_j \right)^2 \geq t \right\} \geq p
\end{eqnarray*}
where $x_j$'s, $t$, and $p$ are arbitrary non-zero constants.
\end{lemma}
\noindent {\it Proof of Lemma~\ref{lem::chi}.}
Without loss of generality, we can assume $x_j> 0$ for all $j$. 
Since $X_j$'s are independent of each other, it suffices to show that  
\begin{eqnarray}
\label{eqn::lem-chi-1}
 \Pr\left\{ \left( a'_j X_j+ \sqrt{J} x_j \right)^2 \geq t \right\} \geq p
\end{eqnarray}
implies 
\begin{eqnarray}
\label{eqn::lem-chi-2}
 \Pr\left\{ \left( a_j X_j+ \sqrt{J} x_j \right)^2 \geq t\right\} \geq p
\end{eqnarray}
for all $j$.
By some algebra,~\eqref{eqn::lem-chi-1} is equivalent to 
\begin{eqnarray}
\label{eqn::lem-chi-3}
\Phi\left( \frac{\sqrt{J}x_j-\sqrt{t}}{a'_j}\right)+\Phi\left( \frac{-\sqrt{J}x_j-\sqrt{t}}{a'_j}\right)\geq p.
\end{eqnarray}
As $J$ goes to infinity, the second term on the left-hand side of~\eqref{eqn::lem-chi-3} goes to $0$. Therefore, we can write~\eqref{eqn::lem-chi-3} as 
\begin{eqnarray}
\label{eqn::lem-chi-4}
\Phi\left( \frac{\sqrt{J}x_j-\sqrt{t}}{a'_j}\right)\geq p.
\end{eqnarray}
Similarly, we can show that~\eqref{eqn::lem-chi-2} is equivalent to 
\begin{eqnarray}
\label{eqn::lem-chi-5}
\Phi\left( \frac{\sqrt{J}x_j-\sqrt{t}}{a_j}\right)\geq p.
\end{eqnarray}
Because $a'_j\geq a_j$,~\eqref{eqn::lem-chi-4} implies~\eqref{eqn::lem-chi-5}. This completes the proof. \QEDB
\medskip

We now prove Theorem~\ref{thm::samplesize-general}. The number of
clusters requires for the test to have power $1-\beta$ should satisfy
\begin{eqnarray*}
\Pr\{J(C\wY)^\top (C\widehat DC^\top)^{-1}(C\wY) \geq \chi^2_{1-\alpha}(k)\mid C\oY =x\} \geq 1-\beta.
\end{eqnarray*}
Theorem~\ref{thm::CLT} implies
\begin{eqnarray*}
\sqrt{J} (C\wY - x) \stackrel{d}{\rightarrow}  \text{N}_k( 0, CDC^\top)
\end{eqnarray*}
Therefore, we can write $C\wY= 1/\sqrt{J} \cdot (CDC^\top)^{1/2} \cdot W_k+x$, where $W_k$ is a $k$-length vector following a standard multivariate normal distribution. As a result, we can write the test statistic as 
\begin{eqnarray*}
\{(CDC^\top)^{1/2} W_k+x\}^\top (C\widehat DC^\top)^{-1} \{(CDC^\top)^{1/2} W_k+x\}
\end{eqnarray*}
By Slutsky's theorem, it has the same asymptotic distribution as
\begin{eqnarray*}
T' &=& \{(CDC^\top)^{1/2} W_k+\sqrt{J}x\}^\top \{C\E(\widehat D)C^\top\}^{-1} \{(CDC^\top)^{1/2} W_k+\sqrt{J}x\}\\
&=&[\{C\E(\widehat D)C^\top\}^{-1/2}(CDC^\top)^{1/2} W_k+\sqrt{J}\{C\E(\widehat D)C^\top\}^{-1/2}x]^\top\\
&&\cdot[ \{C\E(\widehat D)C^\top\}^{-1/2}(CDC^\top)^{1/2} W_k+\sqrt{J}\{C\E(\widehat D)C^\top\}^{-1/2}].
\end{eqnarray*}
From the matrix theory, 
we can write $(CDC^\top)^{1/2} \{C\E(\widehat D)C^\top\}^{-1} (CDC^\top)^{1/2}= P^\top \Lambda P$, where $P$ is an orthogonal matrix and $\Lambda=\text{diag}(\lambda_1,\ldots,\lambda_k)$ is a diagonal matrix.
Because $D_0-D$ is positive semidefinite, $0\leq \lambda_j\leq 1 $ for all $j$.
 Denote $U=(U_1,\ldots, U_m)= P W$, which also follows a standard multivariate normal distribution. Then, we can write  
 \begin{eqnarray*}
T' &=& \left[\Lambda^{1/2}U+ \sqrt{J}\{C\E(\widehat D)C^\top\}^{-1/2}x\right]^\top \left[\Lambda^{1/2}U+ \sqrt{J}\{C\E(\widehat D)C^\top\}^{-1/2}x\right] \\
&=& \sum_{j=1}^k ( \sqrt{ \lambda_j}U_j + \sqrt{J} x'_j)^2,
\end{eqnarray*}
where $x'_j$ is the $j$-th element of $\{C\E(\widehat D)C^\top\}^{-1/2}x$. From Lemma~\ref{lem::chi}, $\Pr(T' \geq t) \geq 1-\beta$ is implied by 
\begin{eqnarray}
\label{eqn::samplesize-general1}
\Pr \left\{\sum_{j=1}^k (U_j + \sqrt{J} x'_j)^2 \geq t \right\} \geq 1-\beta.
\end{eqnarray}
Based on the definition of $s^2( q, 1-\beta,k)$,~\eqref{eqn::samplesize-general1} is equivalent to 
\begin{eqnarray*}
J \sum_{j=1}^k x'^2_j \geq s^2( \chi^2_{1-\alpha}(k), 1-\beta,k).
\end{eqnarray*}
Because $ \sum_{j=1}^k x'^2_j =x^\top \{C\E(\widehat D)C^\top\} x $, we obtain the sample size formula,
\begin{eqnarray*}
J \geq \frac{ s^2( \chi^2_{1-\alpha}(m), 1-\beta,m) }{x^\top \{C\E(\widehat D)C^\top\}^{-1} x }.
\end{eqnarray*}
\QEDB

\subsection{Proof of Theorem~\ref{thm::samplesize-de}}
\label{app::samplesize-de}

We first derive the expression of $\E(\widehat D)$ under
Assumption~\ref{asm::simplification}. From
Appendix~\ref{app::conservative-cov}, we have
\begin{eqnarray*}
\E \{\widehat{\sigma}_{b}^2(1,a)\}
&=& \sigma_{b}^2(1,a) + \frac{1}{J} \sum_{j=1}^J \frac{1}{n_{j1}} \left(1- \frac{n_{j1}}{n_j} \right) \sigma^2_j(1,a)\\
&=& \sigma_b^2 + \frac{1-p_a}{np_a}\sigma_w^2\\
&=&\left\{ r + \frac{(1-p_a)(1-r)}{np_a} \right\} \sigma^2,
\end{eqnarray*}
where the second equality follows from conditions (a),~(b),~and~(c) of
Assumption~\ref{asm::simplification}. Similarly, we obtain
\begin{eqnarray*}
\E \{\widehat{\sigma}_{b}^2(0,a)\}
&=& \sigma_{b}^2(0,a) + \frac{1}{J} \sum_{j=1}^J \frac{1}{n_{j0}} \left(1- \frac{n_{j0}}{n_j} \right) \sigma^2_j(0,a)\\
&=& \sigma_b^2 + \frac{p_a}{n(1-p_a)}\sigma_w^2\\
&=&\left\{ r + \frac{p_a(1-r)}{n(1-p_a)} \right\} \sigma^2
\end{eqnarray*}
and 
\begin{eqnarray*}
\E \{\widehat{\sigma}_{b}^2(1,0;a)\}
&=&\sigma_{b}^2(1,0;a) - \frac{1}{J} \sum_{j=1}^J\frac{\sigma^2_j(1,0;a)}{n_j}\\
&=& \rho \sigma_b^2 - \frac{\rho \sigma_w^2}{n} \\
&=&\rho  \left( r - \frac{1-r}{n}\right)\cdot \sigma^2.
\end{eqnarray*}
Therefore, under Assumption~\ref{asm::simplification},
$\E(\widehat D) = D_0^\ast =\sigma^2\cdot
\text{diag}(D_{01}^\ast,D_{02}^\ast,\ldots, D_{0m}^\ast)$, where
\begin{eqnarray*}
D_{0a}^\ast=\frac{1}{q_a}\begin{pmatrix}
r+\frac{(1-p_a)(1-r)}{np_a} & \rho \left( r- \frac{1-r}{n}\right)\\
 \rho \left( r- \frac{1-r}{n}\right)& r+\frac{p_a(1-r)}{n(1-p_a)}
\end{pmatrix}
\end{eqnarray*}
for $a=1,\ldots,m$.

We next prove the sample size formula. From
Theorem~\ref{thm::samplesize-general}, the number of clusters required
for detecting the alternative hypothesis $H_1^\text{de}: \ADE = x$ with power
$1-\beta$ based on $T_{\text{de}}$ is given as,
\begin{eqnarray*}
J \geq \frac{ s^2( \chi^2_{1-\alpha}(m), 1-\beta,m) }{x^\top \{C_1 \E(\widehat D)C_1^\top\}^{-1} x },
\end{eqnarray*}
which, under Assumption~\ref{asm::simplification}, is equivalent to 
\begin{eqnarray*}
J \geq \frac{ s^2( \chi^2_{1-\alpha}(m), 1-\beta,m)\cdot \sigma^2 }{x^\top \{C_1 D_0^\ast  C_1^\top\}^{-1} x }.
\end{eqnarray*}
Therefore, under the alternative hypothesis $H_1: |\ADE| = \mu$ for all $a$,
the sample size formula is
\begin{eqnarray*}
J &\geq& \frac{ s^2( \chi^2_{1-\alpha}(m), 1-\beta,m)\cdot \sigma^2 }{\mu^2\cdot \bm{1}_m^\top \{C_1 D_0^\ast  C_1^\top\}^{-1} \bm{1}_m }\\
&=& \frac{ s^2( \chi^2_{1-\alpha}(m), 1-\beta,m)\cdot \sigma^2 }{\mu^2 } \cdot \frac{1}{ \sum_{a=1}^m \left\{(1,-1)D_{0a}^\ast(1,-1)^\top\right\}^{-1}}.
\end{eqnarray*}
%
%

Under $r \geq 1/(n+1)$, we have $(1,-1)D_{0a}(1,-1)^\top \geq (1,-1)D_{0a}^\ast(1,-1)^\top$. Thus, a more conservative sample size formula is given as,
\begin{eqnarray*}
J \ \geq \ \frac{ s^2( \chi^2_{1-\alpha}(m), 1-\beta,m)\cdot \sigma^2 }{\mu^2 } \cdot \frac{1}{ \sum_{a=1}^m \left\{(1,-1)D_{0a}(1,-1)^\top\right\}^{-1}}.
\end{eqnarray*}
\QEDB

\subsection{Proof of Theorem~\ref{thm::samplesize-mde}}
\label{app::samplesize-mde}

From Theorem~\ref{thm::samplesize-general}, the number of clusters
required for detecting the alternative hypothesis $H_1^{\text{mde}}: \MDE = \mu$
with power $1-\beta$ based on $T_{\text{mde}}$ is given as,
\begin{eqnarray*}
J \geq \frac{ s^2( \chi^2_{1-\alpha}(m), 1-\beta,m) }{\mu^2 \{C_2 \E(\widehat D)C_2^\top\}^{-1} },
\end{eqnarray*}
which, under Assumption~\ref{asm::simplification}, is equivalent to 
\begin{eqnarray*}
J \geq \frac{ s^2( \chi^2_{1-\alpha}(1), 1-\beta,1)\cdot \sigma^2 }{\mu^2} \sum_{a=1}^m q_a^2\left\{(1,-1)D_{0a}^\ast(1,-1)^\top\right\}.
\end{eqnarray*}
Under $r \geq 1/(n+1)$, we have $(1,-1)D_{0a}(1,-1)^\top \geq (1,-1)D_{0a}^\ast(1,-1)^\top$. Thus, a more conservative sample size formula is given as, 
\begin{eqnarray*}
J \ \geq \ \frac{ s^2( \chi^2_{1-\alpha}(m), 1-\beta,m)\cdot \sigma^2 }{\mu^2 } \cdot \sum_{a=1}^m q_a^2\left\{(1,-1)D_{0a}(1,-1)^\top\right\}.
\end{eqnarray*} \QEDB

\subsection{Proof of Theorem~\ref{thm::samplesize-se}}
\label{app::samplesize-se}

From Theorem~\ref{thm::samplesize-general}, the number of clusters required for detecting the alternative hypothesis $\ASE = x$ with power $1-\beta$ based on $T_{\text{se}}$ is given as,
\begin{eqnarray*}
J \geq \frac{ s^2( \chi^2_{1-\alpha}(2m-2), 1-\beta,2m-2) }{x^\top \{C_3 \E(\widehat D)C_3^\top\}^{-1} x },
\end{eqnarray*}
which, under Assumption~\ref{asm::simplification}, is equivalent to 
\begin{eqnarray*}
J \geq \frac{ s^2( \chi^2_{1-\alpha}(m), 1-\beta,m)\cdot \sigma^2 }{x^\top \{C_3 D_0^\ast  C_3^\top\}^{-1} x }.
\end{eqnarray*}
Therefore, under the alternative hypothesis $H_1^{\text{se}}: \max_{a\neq a'}|\ASE(z;a,a')| = \mu$ for $z=0,1$, the sample size formula is 
\begin{eqnarray*}
J \geq \frac{ s^2( \chi^2_{1-\alpha}(2m-2), 1-\beta,2m-2)\cdot \sigma^2 }{\mu^2 \cdot \min_{s\in \mathcal{S}} s^\top \{C_3 D_0^\ast C_3^\top\}^{-1} s },
\end{eqnarray*}
where $\mathcal{S}$ is the set of $s = (\ASE(0;1,2),\ASE(0;2,3),\ldots, \ASE(0;m-1,m),\ASE(1;1,2),\ASE(1;2,3),\ldots, \allowbreak\ASE(1;m-1,m))$ satisfying $\max_{z,a\neq a'} |\ASE(a,a')| = 1$ for $z=0,1$. \QEDB

\subsection{Proof of Theorem~\ref{thm::reg}}
\label{app::proof-thm9}
We first prove the equivalence between the point estimators.
The OLS estimate can be written as,
\begin{eqnarray*}
 \widehat{\bm{\beta}} \ = \  ( \bX^\top \bW \bX) ^{-1}  \bX^\top \bW \bY.
\end{eqnarray*}
Because the columns of $\bX$ are orthogonal to each other, we can consider each element of  $\widehat{\bm{\beta}}$ separately.
Therefore, we have 
\begin{eqnarray*}
 \widehat{\bm{\beta}}_{za} &=&  \left\{\sum_{j=1}^J \sum_{i=1}^{n_j} \bone(Z_{ij}=z, A_{j}=a) w_{ij}\right\}^{-1} \left\{\sum_{j=1}^J \sum_{i=1}^{n_j}  \bone(Z_{ij}=z, A_{j}=a) w_{ij}Y_{ij}\right\}\\
 &=& \sum_{j=1}^J\sum_{i=1}^{n_j} \frac{1}{J_a n_{jz}}\cdot  \bone(Z_{ij}=z, A_{j}=a) Y_{ij}\\
 &=& \wY(z,a).
\end{eqnarray*}

We then prove the equivalence between the variance estimators. Recall the variance estimator,
\begin{eqnarray*}
\widehat{\var}^{\textnormal{cluster}}_{\textnormal{hc2}}(\widehat{\bm{\beta}})\ =\ (\bX^\top \bW \bX)^{-1} \left\{ \sum_j \bX_j^\top \bW_j (\bI_{n_j}-\bP_j)^{-1/2} \widehat{\bepsilon}_j \widehat{\bepsilon}_j^\top (\bI_{n_j}-\bP_j)^{-1/2} \bW_j \bX_j \right\} (\bX^\top \bW \bX)^{-1},
\end{eqnarray*}
where $\bI_{n_j}$ is the $n_j \times n_j$ identity matrix and $\bP_j$
is the following cluster leverage matrix,
\begin{eqnarray*}
 \bP_j \ = \  \bW_j^{1/2} \bX_j(\bX^\top \bW \bX)^{-1}
 \bX_j^\top \bW_j^{1/2}.
\end{eqnarray*}

Without loss of generality, suppose $A_j=1$. We have 
\begin{eqnarray*}
 (\bX^\top \bW \bX)^{-1} &=& \bm{I}_{n\times n},\\
	\bP_j & = & \bW_j^{1/2}  \bX_j(\bX^\top \bW \bX)^{-1}
	\bX_j^\top \bW_j^{1/2} \\
	&=&   \begin{pmatrix}
		\frac{1}{\sqrt{ J_1 n_{j1}}} \bm{1}_{n_{j1}} & \bm{0}_{n_{j1}}  \\
		 \bm{0}_{n_{j0}} & \frac{1}{\sqrt{ J_1 n_{j0}}} \bm{1}_{n_{j0}}
	\end{pmatrix}   \begin{pmatrix}
	\frac{1}{\sqrt{ J_1 n_{j1}}} \bm{1}_{n_{j1}} & \bm{0}_{n_{j1}}  \\
	\bm{0}_{n_{j0}} & \frac{1}{\sqrt{ J_1 n_{j0}}} \bm{1}_{n_{j0}}
\end{pmatrix}^\top\\
	&=& \begin{pmatrix}
		\frac{1}{J_a n_{j1}} \bm{1}_{n_{j1} \times n_{j1} }& \bm{0}_{n_{j1} \times n_{j0} } \\
		\bm{0}_{n_{j0} \times n_{j1} } &\frac{1}{J_a n_{j0}} \bm{1}_{n_{j0} \times n_{j0}} 
	\end{pmatrix},
\end{eqnarray*} 
where $\bm{I}_k $ is an $k$-dimensional identity matrix, $\bm{1}_k$ ($\bm{0}_k$) is an $k$-dimensional vector of ones
(zeros) and $\bm{1}_{k_1\times k_2}$ ($\bm{0}_{k_1\times k_2}$) is an
$k_1 \times k_2$ dimensional matrix of ones (zeros).

Since $( \bm{1}_{n_{j1}}^\top, \bm{0}_{n_{j0}}^\top )^\top$ and
$( \bm{0}_{n_{j1}}^\top, \bm{1}_{n_{j0}}^\top )^\top$ are two
eigenvectors of $\bm{I}_{n_j}-\bP_j$ whose eigenvalue is
$(J_1-1)/J_1$, we have,
\begin{eqnarray*}
	(\bm{I}_{n_j}-\bP_j)^{-1/2} ( \bm{1}_{n_{j1}}^\top,
	\bm{0}_{n_{j0}}^\top )^\top & = &  \sqrt{\frac{J_1}{J_1-1}} ( \bm{1}_{n_{j1}}^\top, \bm{0}_{n_{j0}}^\top )^\top,\\
	(\bm{I}_{n_j}-\bP_j)^{-1/2} ( \bm{0}_{n_{j1}}^\top, \bm{1}_{n_{j0}}^\top
	)^\top & = &  \sqrt{\frac{J_1}{J_1-1}} ( \bm{0}_{n_{j1}}^\top, \bm{1}_{n_{j0}}^\top )^\top.
\end{eqnarray*}
Thus, 
\begin{eqnarray*}
	(\bm{I}_{n_j}-\bP_j)^{-1/2} \bW_j \bX_j & = & \sqrt{\frac{J_1}{J_1-1}}  \begin{pmatrix}
		\frac{1}{J_1 n_{j1}} \bm{1}_{n_{j1}}& \bm{0}_{n_{j1}}& \bm{0}_{n_{j1}\times (2m-2)} \\
		\bm{0}_{n_{j0}} & \frac{1}{ J_1 n_{j0}} \bm{1}_{n_{j0}} & \bm{0}_{n_{j0}\times (2m-2)}
	\end{pmatrix}.
\end{eqnarray*}

For a unit with $(A_j=1,Z_{ij}=1)$, we have
$\widehat{\epsilon}_{ij} = Y_{ij}-
\widehat{\beta}_{11}= Y_{ij}- \wY(1,1)$,
and for a unit with $(A_j=1,Z_{ij}=0)$, we have
$\widehat{\epsilon}_{ij} = Y_{ij}- \widehat{\alpha}_{01}= Y_{ij}-
\wY(0,1)$. As a result,
\begin{eqnarray*}
	&&\widehat{\bepsilon}_j^\top (\bm{I}_{n_j}-\bP_j)^{-1/2} \bW_j \bX_j \\
	&=& \sqrt{\frac{J_1}{J_1-1}}
	( Y_{1j}- \widehat{Y}(1,1), \ldots, Y_{n_{j}j}- \widehat{Y}(0,1))
	 \begin{pmatrix}
		\frac{1}{J_1 n_{j1}} \bm{1}_{n_{j1}}& \bm{0}_{n_{j1}}& \bm{0}_{n_{j1}\times (2m-2)} \\
		\bm{0}_{n_{j0}} & \frac{1}{ J_1 n_{j0}} \bm{1}_{n_{j0}} & \bm{0}_{n_{j0}\times (2m-2)}
	\end{pmatrix},\\
	&=&\sqrt{\frac{J_1 }{J_1-1}} \begin{pmatrix}
		\frac{1}{ J_1 n_{j1}}  \left\{ \sum_{i=1}^{n_j} Y_{ij}Z_{ij} - n_{j1} \widehat{Y}(1,1) \right\}\\
	\frac{1}{J_1 n_{j0}}  \left\{ \sum_{i=1}^{n_j} Y_{ij}(1-Z_{ij}) - n_{j0} \widehat{Y}(0,1) \right\}
	\end{pmatrix} ^\top\\
	&=& \sqrt{\frac{1}{J_1(J_1-1)}} 
   \left( \widehat{Y}_j(1)-\widehat{Y}(1,1), \widehat{Y}_j(0)-\widehat{Y}(0,1),\bm{0}^\top_{2m-2}\right).
\end{eqnarray*}
Similar result applies for $A_j=a$, where $a=1.\ldots,J$.
Therefore, $\widehat{\var}^{\textnormal{cluster}}_{\textnormal{hc2}}(\widehat{\bm{\beta}})$ is a block diagonal matrix with the $a$-th block
{\small
\begin{eqnarray*}
&&\frac{1}{J_a(J_a-1)}\sum_{j=1}^J \bone(A_j=a)  \left( \widehat{Y}_j(1)-\widehat{Y}(1,a), \widehat{Y}_j(0)-\widehat{Y}(0,a)\right)\left( \widehat{Y}_j(1)-\widehat{Y}(1,a), \widehat{Y}_j(0)-\widehat{Y}(0,a)\right)^\top\\
&=&\begin{pmatrix}
\frac{\sum_{i=1}^{J} \left \{ \wY_j(1)- \wY(1,a)\right \}^2\bone(A_j=a)}{J_a(J_a-1)} &\frac{\sum_{i=1}^{J} \left \{ \wY_j(1,a)- \wY(1,a)\right \}\left \{ \wY_j(0,a)- \wY(0,a)\right \}\bone(A_j=a)}{J_a(J_a-1)} \\
\frac{\sum_{i=1}^{J} \left \{ \wY_j(1,a)- \wY(1,a)\right \}\left \{ \wY_j(0,a)- \wY(0,a)\right \}\bone(A_j=a)}{J_a(J_a-1)} & \frac{\sum_{i=1}^{J} \left \{ \wY_j(0)- \wY(0,a)\right \}^2\bone(A_j=a)}{J_a(J_a-1)}
\end{pmatrix}\\
&=& \frac{\widehat{D}}{J}.
\end{eqnarray*}
} \QEDB

\subsection{Proof of Theorem~\ref{thm::comparison}}
\label{app:comparison}

First, we calculate the variance of $\widehat \ATE$ under the two-stage randomized design.
 In this case, $\widehat{\ATE}$ is the same as $\widehat{\ADE}$. From Theorem~\ref{thm::var-general}, we have 
 \begin{eqnarray*}
\textnormal{var} \left(\widehat{\ADE}\right)&=&\sum_{a=1}^{m} \frac{J_a^2}{J^2}\cdot \textnormal{var} \left\{\widehat{\ADE}(a)\right\}+\sum_{a \neq a'} \frac{J_aJ_{a'}}{J^2} \cdot \textnormal{cov} \left\{\widehat{\ADE}(a),\widehat{\ADE}(a')\right\}.
\end{eqnarray*}
When there is no interference, we have 
{\small
\begin{eqnarray*}
&&\var \left\{ \widehat{\mathsf{DEY}}(a)\right\} \\
&=& \left(1- \frac{J_a}{J} \right) \frac{\tau_b^2 }{J_a}
+\frac{1}{J_a J} \sum_{j=1}^J\left\{ \frac{\sum_{i=1}^n(Y_{ij}(1)-\oY_j(1))^2}{ (n-1) np_a}+ \frac{\sum_{i=1}^n(Y_{ij}(0)-\oY_j(0))^2}{ (n-1) n(1-p_a)}- \frac{\sum_{i=1}^n(\ATE_{ij}-\ATE_j)^2}{ (n-1) n}\right\}\\
&=& \left(1- \frac{J_a}{J} \right) \frac{\tau_b^2 }{J_a} +\frac{nJ-1}{(n-1)nJ_a J} \left\{\frac{\eta^2_w(1)}{p_a} +\frac{\eta^2_w(0)}{1-p_a}-\tau_w^2 \right\}
\end{eqnarray*}
}
and $ \textnormal{cov} \left\{\widehat{\ADE}(a),\widehat{\ADE}(a')\right\}=- \tau_b^2/J$.
Therefore, we can obtain
\begin{eqnarray*}
&&\textnormal{var} \left(\widehat{\ADE}\right) \\
&=& \sum_{a=1}^m J_a \left(1- \frac{J_a}{J} \right) \frac{\tau_b^2 }{J^2} -\sum_{a \neq a'}^m \frac{J_aJ_{a'}}{J^2} \cdot \frac{\tau^2_b}{J}+ \sum_{a=1}^{m} \frac{J_a^2}{J^2} \cdot \frac{nJ-1}{(n-1)nJ_a J} \left\{\frac{\eta^2_w(1)}{p_a} +\frac{\eta^2_w(0)}{1-p_a}-\tau_w^2 \right\}\\
&=& \frac{nJ-1}{J^3(n-1)} \sum_{a=1}^m \frac{J_a }{np_a} \cdot \eta^2_w(1) + \frac{nJ-1}{J^3(n-1)} \sum_{a=1}^m \frac{J_a }{n(1-p_a)} \cdot \eta_w^2(0)- \frac{nJ-1}{J^3(n-1)} \sum_{a=1}^m \frac{J_a }{n} \cdot \tau_w^2\\
&=& \frac{(nJ-1)(1-r)}{nJ^3} \left\{\sum_{a=1}^m \frac{J_a }{np_a} \cdot \eta^2(1) + \sum_{a=1}^m \frac{J_a }{n(1-p_a)} \cdot \eta^2(0)- \sum_{a=1}^m \frac{J_a }{n} \cdot \tau^2\right\}\\
&\approx& \frac{1-r}{J^2} \sum_{a=1}^m \frac{J_a }{np_a} \cdot \eta^2(1) + \frac{1-r}{J^2} \sum_{a=1}^m \frac{J_a }{n(1-p_a)} \cdot \eta^2(0)- \frac{1-r}{nJ}  \cdot \tau^2,
\end{eqnarray*}
where the last line follows from the approximation assumptions in equation~\eqref{eqn::compare-approx}.

Second, the variance of $\widehat \ATE$ under the completely randomized experiment with the number of the treated units equal to $ \sum_{a=1}^m J_an p_a$ is given as,
\begin{eqnarray*}
&&\frac{1}{\sum_{a=1}^m J_an p_a} \cdot \eta^2(1)+\frac{1}{\sum_{a=1}^m J_an (1-p_a)} \cdot \eta^2(0)-\frac{1}{Jn} \cdot \tau^2.
\end{eqnarray*}

Third, we calculate the variance of $\widehat \ATE$ under cluster randomized experiments with the same number of treated units. 
In the cluster randomized experiments, the units in each cluster get the same treatment condition. Thus, the number of the treated clusters is $\sum_{a=1}^m J_ap_a$. As a result, the variance of $\widehat \ATE$ is given as,
\begin{eqnarray*}
&& \frac{\eta_b^2(1)}{\sum_{a=1}^m J_ap_a}+\frac{\eta_b^2(0)}{\sum_{a=1}^m J_a(1-p_a)} - \frac{ \tau_b^2}{J} \cdot\\
 &\approx&\frac{1+(n-1)r}{\sum_{a=1}^m J_anp_a} \cdot \eta^2(1)+\frac{1+(n-1)r}{\sum_{a=1}^m J_an(1-p_a)} \cdot \eta^2(0)- \frac{1+(n-1)r}{nJ} \cdot \tau^2,
\end{eqnarray*}
where the last line follows from the approximation assumptions in equation~\eqref{eqn::compare-approx}.
\QEDB

\section{Computational details}
\label{app:computation}
 
 We provide a strategy for numerically calculating the required number of clusters in Theorem~\ref{thm::samplesize-se}.
 We focus on the following optimization problem,
 $$\min_{s\in \mathcal{S}} s^\top \{C_3 D_0C_3^\top\}^{-1} s,$$ where $a = (\ASE(0;1,2),\ASE(0;2,3),\ldots, \ASE(0;m-1,m),\ASE(1;1,2),\ASE(1;2,3),\ldots, \ASE(1;m-1,m))$ satisfies the constraint $\max_{a\neq a'} |\ASE(z;a,a')| = 1$ for $z=0,1$.

We consider all the possible cases in which $\max_{a\neq a'} |\ASE(z;a,a')| = 1$  holds for $z=0,1$. First,
using quadratic programming, we can obtain the minimum of $s^\top \{C_3 D_0C_3^\top\}^{-1} s$ under the constraint $\ASE(1;1,2) = 1$, $\ASE(0;1,2) = 1$ and   $ -1 \leq \ASE(z;a,a') \leq 1$ for all $z,a,a'$. We denote it by $l(1,2;1,2)$. 
Similarly, we can obtain $l(a_1,a_1'; a_0,a_0' )$ for all $a_1,a_1',a_0,a_0'$ by implementing this procedure for each of the possible cases satisfying $\max_{a\neq a'} |\ASE(z;a,a')| = 1$ for $z=0,1$. As a result, the solution to the optimization problem is $\min l(a_1,a_1'; a_0,a_0' )$.

\section{Simulation Studies}
\label{sec::simulation}

We conduct simulation studies to evaluate the empirical performance of
the sample size formulas for the direct, marginal direct, and
spillover effects. We consider a two-stage randomized experiment with
three different treatment assignment mechanisms ($m=4$), under which
the treated proportions are $20\%$, $40\%$, $60\%$, and $80\%$,
respectively. We generate the treatment assignment mechanism $A_j$
with $\Pr(A_j = a) = 1/4$ for $a=1,2,3,4$ such that $J_a=J/4$. We then
completely randomize the treatment assignment $Z_{ij}$ within each
cluster according to the selected assignment mechanism.

Our data generating process is as follows. First, we generate the
cluster-level average potential outcomes as,
\begin{eqnarray*}
\oY_{j}(0,a) \sim \text{N}(\theta_{0a}, \sigma^2_b), \quad \oY_j(1,a)& \sim& \text{N}( \theta_{1a}+ \rho \{\oY_{j}(0,a)-\theta_{0a}\} , (1-\rho^2) \sigma^2_b)
\end{eqnarray*}
for $a=1,2,3,4$. Second, we generate the individual-level average
potential outcomes $Y_{ij}(z,a)$ as,
 \begin{eqnarray*}
\begin{pmatrix} Y_{ij}(1,a) \\ Y_{ij}(0,a) \end{pmatrix} \sim \text{N}_2\left( \begin{pmatrix} \oY_j(1,a) \\ \oY_j(0,a) \end{pmatrix}, \begin{pmatrix} \sigma^2_w & \rho \sigma_w^2\\ \rho \sigma_w^2 & \sigma_w^2\end{pmatrix}\right)
\end{eqnarray*}
for $a=1,2,3,4$. In this super population setting, the direct effect
under treatment assignment mechanism $a$ is given by
$\theta_{1a}-\theta_{0a}$ for $a=1,2,3,4$, whereas the marginal direct
effect equals
$(\theta_{11}+\theta_{12}+\theta_{13}+\theta_{14})/4-(\theta_{01}+\theta_{02}+\theta_{03}+\theta_{04})/4$.
The spillover effect comparing treatment assignment mechanisms $a$ and
$a'$ under treatment condition $z$ is $ \theta_{za}-\theta_{za'}$ for
$z=0,1$ and $a,a'=1,2,3,4$. However, our target causal quantities of interest are
finite-sample causal effects ($\ADE(a)$, $\MDE$, $\ASE(z;a,a')$),
which generally do not equal their super-population counterparts due
to sample variation. Therefore, we center the generated potential
outcomes so that the finite-sample and super-population causal effects
are equal to one another, i.e., $\oY(z,a) = \theta_{za}$ for $z=0,1$
and $a=1,2,3,4$.

We choose different values of $\theta$'s based on the different
alternative hypotheses for our three causal effects of interest. For
the direct effect, we generate $\theta_{0a}$ ($a=1,2,3,4$) from a
uniform distribution on the interval $[-0.3,0.3]$, and set
$\theta_{1a}=0.3+\theta_{0a}$ for all $a$; the generated potential outcomes
satisfy $ |\ADE(a)|=0.3$ for all $a$. For the marginal direct effect, we
generate $\theta_{0a}$ ($a=1,2,3,4$) from a uniform distribution on the
interval $[-0.3,0.3]$ and set $\theta_{11}=0.12+\theta_{01}$,
$\theta_{12}=0.48+\theta_{02}$, $\theta_{13}=0.24+\theta_{03}$, and $\theta_{14}=0.36+\theta_{04}$; the
generated potential outcomes satisfy $\MDE=0.3$. For the spillover
effect, we generate $\theta_{0a}$ and $\theta_{1a}$
 from a uniform distribution on the interval $[-0.15,0.15]$ for  ($a=1,2, 3$) 
and set $\theta_{z4} = 0.3+ \min(\theta_{z1},\theta_{z2},\theta_{z3})$ for $z=0,1$; the
generated potential outcomes satisfy
$\max_{a\neq a'}|\ASE(Z;a,a')|=0.3$ for $z=0,1$.

We first consider the scenario with equal cluster size $n$ for all
clusters, the total variance $\sigma^2=1$, and two levels of cluster
size ($n=20$ and $n=100$).  We choose three values of the
  correlation coefficient between potential outcomes,
  $\rho=0,0.3,0.6$. Because the sample size formulas in
  equations~\eqref{eqn::samplesize-de3},~\eqref{eqn::samplesize-mde3},
  and~\eqref{eqn::samplesize-se3} assume $\rho=0$, the simulation
  settings with $\rho=0.3,0.6$ evaluate their robustness to the
  misspecification of this design parameters. In each setting, we
vary the intracluster correlation coefficient
$r=\sigma^2_b/(\sigma^2_w+\sigma_b^2)$ from $0$ to $1$, which also
determines the values of $\sigma_w^2$ and $\sigma_b^2$. We compute the
required number of clusters using the sample size formulas and then
generate the data based on the resulting number of clusters. The
statistical power is estimated under each setting by averaging over
$1,000$ Monte Carlo simulations.

\begin{figure}[t!]
 \centering
 \includegraphics[width=\textwidth]{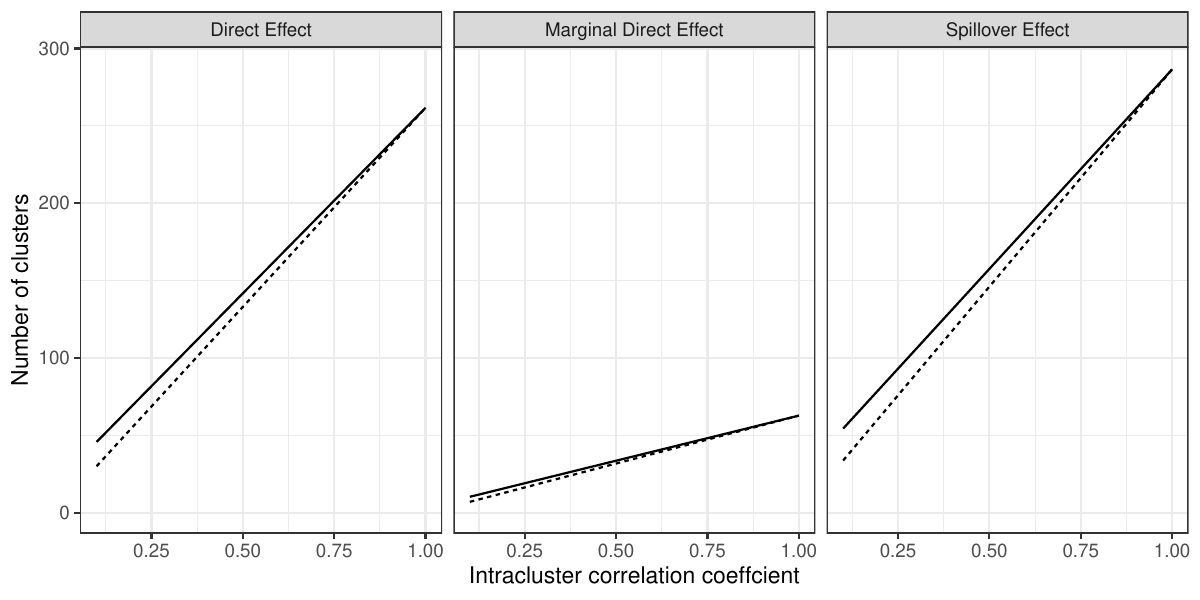}
 \caption{The required number of clusters calculated from
 equations~\eqref{eqn::samplesize-de3},~\eqref{eqn::samplesize-mde3},
 and~\eqref{eqn::samplesize-se3} for the statistical power of
 80\%. The parameters are set to $\sigma^2=1$, $\mu=0.3$,
 $\alpha=0.05$, $\beta=0.2$ with the intracluster correlation
 coefficient varying from $0$ to $1$ (horizontal axis). The solid
 lines indicate the setting with cluster size of $n=20$, and the
 dashed lines indicate the setting with $n=100$.}
\label{fig::samplesize}
\end{figure}

Figure~\ref{fig::samplesize} shows the required number of clusters
calculated from
equations~\eqref{eqn::samplesize-de3},~\eqref{eqn::samplesize-mde3},
and~\eqref{eqn::samplesize-se3} for the statistical power of 80\%.
The parameters are set to $\sigma^2=1$, $\mu=0.3$, $\alpha=0.05$, and
$\beta=0.2$ with the intracluster correlation coefficient varying from
$0$ to $1$ (horizontal axis). The required number of clusters for the
marginal direct effect (middle panel) is much less than those for the
direct and spillover effects (left and right panels, respectively).
Across all settings, the required cluster number increases linearly
with the intracluster correlation coefficient. The difference between
the settings with a small cluster size $n=20$ and a moderate cluster size
$n=100$ is not substantial. This is because the conservative variance
(covariance) matrix estimators rely solely on the estimated
between-cluster variances, in which the cluster size plays a minimal
role. As a result, having a large cluster size does not affect the
required number of clusters significantly.

\begin{figure}[p]
\centering  
 \subfigure[Equal cluster
 size]{\label{fig::power1}\includegraphics[width=1\textwidth]{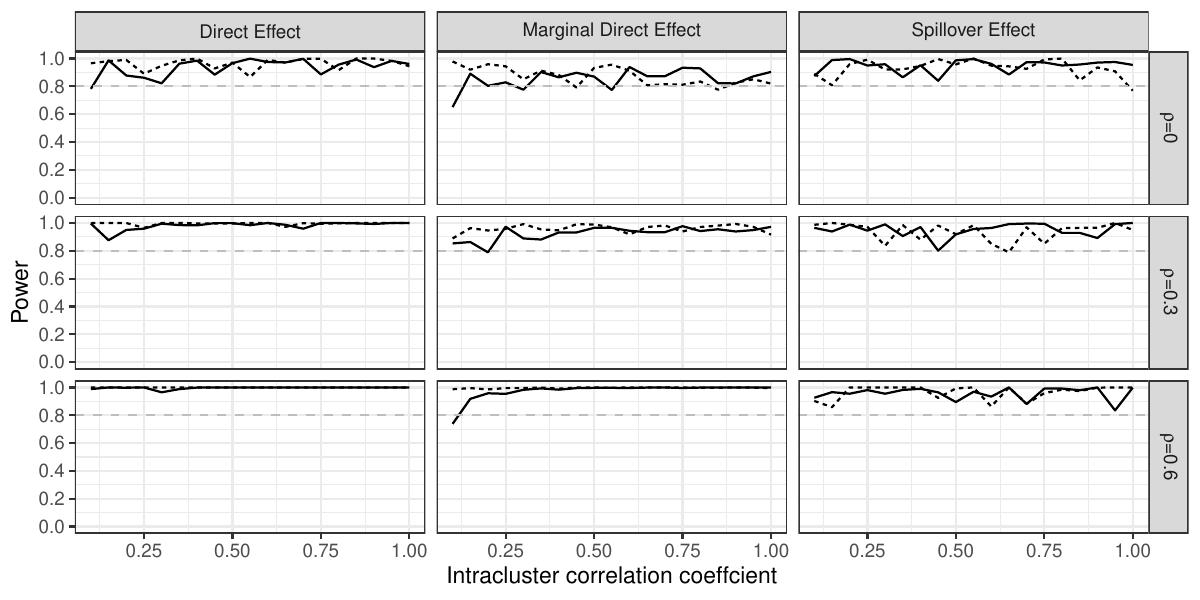}}
 \subfigure[Unequal cluster size]{\label{fig::power2}\includegraphics[width=1\textwidth]{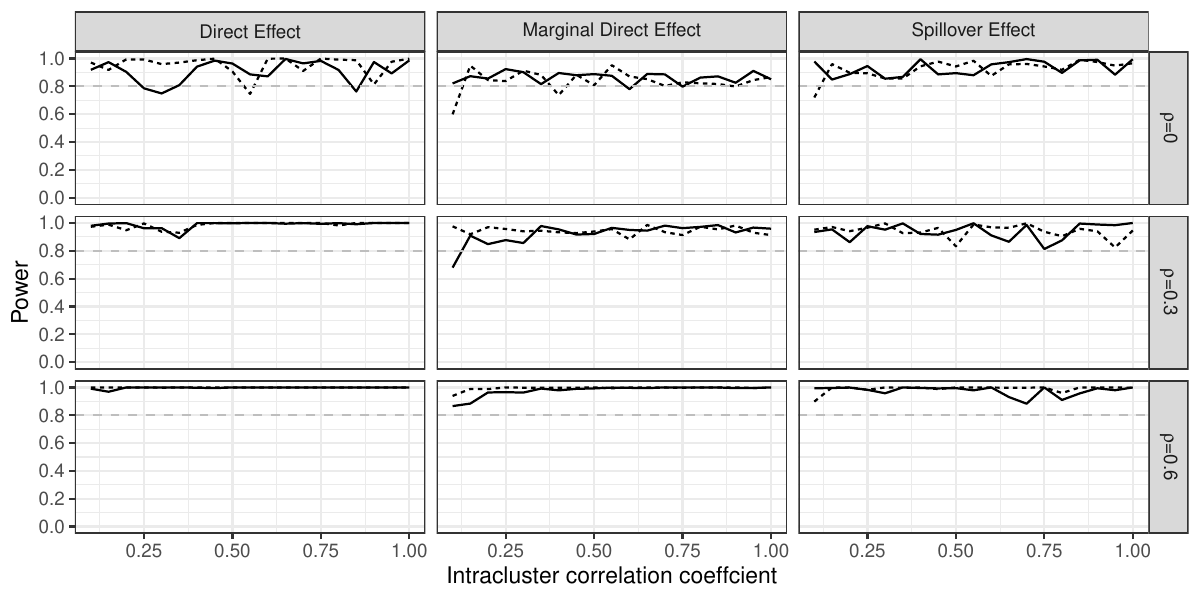}}
 \caption{Estimated statistical power for testing the alternative
 hypotheses the direct, marginal direct, and spillover effects. The
 solid lines indicate the setting with cluster size of $n=20$, and
 the dashed lines indicate the setting with $n=100$. In each plot,
 we vary the correlation between potential outcomes $\rho$ as well
 as the intracluster correlation coefficient (horizontal axis). }
\label{fig::power}
\end{figure}

Figure~\ref{fig::power1} presents the estimated statistical power for
testing the alternative hypotheses concerning the direct, marginal
direct, and spillover effects in the left, middle, and right plots,
respectively.  With the correct specification of the correlation
  coefficient $\rho$, the achieved power is close to its expected
  level ($0.8$) under almost all settings for the direct effect,
  marginal direct effect, and spillover effect.  When the intracluster
  correlation coefficient is small, the statistical power for the
  direct effect and marginal direct effect is sometimes below the
  nominal level of $0.8$. This may arise because the required number
  of clusters is small under these settings (e.g., $J\ge 20$ for the
  marginal direct effect when the intracluster correlation coefficient
  is $0.1$), reducing the accuracy of the asymptotic approximation
  used by the sample size formulas.

  With the misspecified values of correlation coefficient
  $\rho=0.3, 0.6$, the power is close to $1$ for the direct effect and
  marginal direct effect when the intracluster correlation coefficient
  is moderate or large. This suggests that the sample size formula is
  conservative for these quantities.  In contrast, the power is
  smaller than the expected level for the spillover effect, especially
  with a large value of the intracluster correlation coefficient. This
  suggests that the sample size formula for the spillover effect may
  not be robust to the misspecification of the correlation
  coefficient.

  Next, we consider the scenario with unequal cluster size. We
  generate each cluster size from a categorical distribution
  distribution taking values in $\{0.5n,0.75n,n,1.5n,2n,2.5n\}$ with
  probabilities $\{0.25,0.1,0.1,0.1,0.2,0.25\}$, respectively. We then
  generate the data using the number of clusters calculated from the
  sample size formulas. The parameter $\bar{n}$ in these formulas is
  calculated based on the distribution of the cluster sizes. Other
  parameters are the same as those of the case with equal cluster
  size.

  Figure~\ref{fig::power2} shows the results.  The results for the
  direct effect and marginal direct effect are largely similar to
  those presented in Figure~\ref{fig::power1}.  For the spillover
  effect, the variation of power is larger with unequal cluster size
  than with equal cluster size when the intracluster correlation
  coefficient is misspecified.  These results show that the sample
  size formulas are robust to the unequal cluster sizes.

  The simulation results also suggest that the sample size formulas
  are robust to the violation of the simplifying conditions used in
  Assumption~\ref{asm::simplification}. The reason is that the
  variances in the generated data do not satisfy these simplifying
  conditions due to finite sample variation.

\end{document}